\documentclass[journal=jacsat,manuscript=article]{achemso}

\usepackage{tikz}
\usepackage[version=3]{mhchem}

\usepackage{chemformula} 

\usepackage{graphicx}%
\usepackage{multirow}%
\usepackage{amsmath,amssymb,amsfonts}%
\usepackage{amsthm}%
\usepackage{mathrsfs}%
\usepackage[title]{appendix}%
\usepackage{xcolor}%
\usepackage{textcomp}%
\usepackage{manyfoot}%
\usepackage{booktabs}
\usepackage{array}
\usepackage{caption}
\usepackage{algorithm}%
\usepackage{algorithmicx}%
\usepackage{algpseudocode}%
\usepackage{listings}%
\usepackage{soul}
\usepackage{float}
\usepackage{lineno}
\usepackage{pifont}

\soulregister{\cite}7
\soulregister{\citep}7
\soulregister{\citet}7
\soulregister{\ref}7
\soulregister{\pageref}7
\soulregister{\textit}7

\newcommand{\subsubsubsection}[1]{
    \vspace{0.5em}
    \noindent\textbf{#1}
    \vspace{0.2em}
}

\title{Bridging Visual Intuition and Chemical Expertise: An Autonomous Analysis Framework for Nonadiabatic Dynamics Simulations via \textit{Mentor-Engineer-Student} Collaboration}
\author{Yifei Zhu}
\affiliation{SCNU Environmental Research Institute, Guangdong Provincial Key Laboratory of Chemical Pollution and Environmental Safety, School of Environment, MOE Key Laboratory of Environmental Theoretical Chemistry, South China Normal University, Guangzhou 510006, P. R. China.}
\altaffiliation{These authors contributed equally.}
\author{Jiahui Zhang}
\affiliation{SCNU Environmental Research Institute, Guangdong Provincial Key Laboratory of Chemical Pollution and Environmental Safety, School of Environment, MOE Key Laboratory of Environmental Theoretical Chemistry, South China Normal University, Guangzhou 510006, P. R. China.}
\altaffiliation{These authors contributed equally.}
\author{Binni Huang}
\affiliation{SCNU Environmental Research Institute, Guangdong Provincial Key Laboratory of Chemical Pollution and Environmental Safety, School of Environment, MOE Key Laboratory of Environmental Theoretical Chemistry, South China Normal University, Guangzhou 510006, P. R. China.}
\author{Zhenggang Lan}
\affiliation{SCNU Environmental Research Institute, Guangdong Provincial Key Laboratory of Chemical Pollution and Environmental Safety, School of Environment, MOE Key Laboratory of Environmental Theoretical Chemistry, South China Normal University, Guangzhou 510006, P. R. China.}
\email{zhenggang.lan@m.scnu.edu.cn}

\begin{document}
\graphicspath{{Figures/}}
\clearpage
\abstract{
Analyzing nonadiabatic molecular dynamics trajectories traditionally heavily relies on expert intuition and visual pattern recognition, a process that is difficult to formalize.
We present VisU, a vision-driven framework that leverages the complementary strengths of two state-of-the-art large language models (Doubao-Seed-1.6-Vision and DeepSeek-V3.2) to establish a ``virtual research cooperation.''
This operates through a ``\textit{Mentor-Engineer-Student}'' paradigm that mimics the collaborative intelligence of a professional chemistry laboratory.
Within this ecosystem, the \textit{Mentor} provides physical intuition through visual reasoning, while the \textit{Engineer} adaptively constructs analysis scripts, and the \textit{Student} executes the pipeline and managing the data and results.
VisU autonomously orchestrates a four-stage workflow comprising \textit{Preprocessing}, \textit{Recursive Channel Discovery}, \textit{Important-Motion Identification}, and \textit{Validation/Summary}.
This systematic approach identifies reaction channels and key nuclear motions while generating a professional academic report at the end.
By bridging visual insight with chemical expertise, VisU establishes a new paradigm for human-AI collaboration in the analysis of excited-state dynamics
simulation results, significantly reducing dependence on manual interpretation and enabling more intuitive, scalable mechanistic discovery.
}
\clearpage
\maketitle

\section{Introduction}

Nonadiabatic dynamics plays a central role in photophysics, photochemistry, and photobiology, \cite{domcke2012role, matsika2011nonadiabatic,domcke2004conical, domcke2011conical} arising from the breakdown of the Born-Oppenheimer approximation due to strong electron-nuclear coupling. \cite{domcke2004conical, domcke2011conical, matsika2021electronic}
For polyatomic systems, the rapid growth of nuclear degrees of freedom makes trajectory-based nonadiabatic molecular dynamics (NAMD) a widely adopted tool. \cite{domcke2004conical, domcke2011conical, matsika2021electronic, crespo2018recent, Gonzalez2020book, curchod2018ab}
It effectively balances computational cost with the ability to capture complex photoreaction mechanisms. \cite{mai2018nonadiabatic, crespo2018recent, akimov2013theoretical, du2015fly, tapavicza2007trajectory, nelson2014nonadiabatic}
However, the analysis of high-dimensional trajectories, particularly in multichannel scenarios, presents a fundamental bottleneck. \cite{tully2012perspective,zhuUnsupervisedMachineLearning2024}
The composite nature of molecular motions makes subtle branching behaviors difficult to identify without intensive, expert-guided interpretation.

Distilling chemically meaningful insights from the output data obtained from the trajectory-based NAMD simulations  requires the integration of visual reasoning with domain-specific knowledge.
Researchers routinely inspect trajectory evolutions, critical structures and dynamical animations to correlate molecular spatial patterns with photochemical concepts, such as conical intersections and reaction coordinates
This tightly-coupled visual-chemical reasoning loop is essential for distinguishing nonadiabatic channels and identifying governing nuclear motions.

Recent advances in unsupervised machine learning (ML) provided valuable tools for dimensionality reduction and clustering to handle high-dimensional trajectory data. \cite{zhuUnsupervisedMachineLearning2024,zhu2022analysis,glielmo2021unsupervised, ceriottiUnsupervisedMachineLearning2019,virshup2012nonlinear, achesonAutomaticClusteringExcitedState2023, belyaev2015nonadiabatic, capano2017photophysics, delmasAutomatedSelectionNuclear2025,how2021significance, how2022dimensionality, karaDONKEYFlexibleAccurate2025, kochmanNonadiabaticMolecularDynamics2024, liAnalysisGeometricalEvolution2017, liAnalysisTrajectorySimilarity2018, linPredictionExcitedstateReaction2021, mangan2021dependence, pengAnalysisBathMotion2021a, richings2021analyzing, tavadze2018machine, zhou2020structural, liu2025automatedframeworkanalyzingstructural,tsutsumiReactionSpaceProjector2022, tsutsumiVisualizationIntrinsicReaction2018}
However, these purely data-driven methods typically lack the interpretive capacity of a chemist, who perceives data patterns through a combination of visual intuition and domain-specific knowledge.
Consequently, advanced automated pipelines \cite{zhuPrincipalComponentAnalysis2022a,zhuUnsupervisedMachineLearning2024,pinheiroULaMDynEnhancingExcitedstate2025,liu2025automatedframeworkanalyzingstructural} still require continuous human intervention for visual validation and parameter tuning. \cite{zhuUnsupervisedMachineLearning2024}
In one word, the human experts still play essential roles in such analysis protocols, due to their visiual preception and professional knowledge.


The rise of large language models (LLMs) offers new opportunities for orchestrating scientific workflows across various scientific research fields. \cite{ai4science2023impactlargelanguagemodels, jimenez2024swebench,laurent2024labbenchmeasuringcapabilitieslanguage,miret2025enabling,white2023future, jablonka202314,ramos2025review,chenSciToolAgentKnowledgeGraphDriven2025}
However, conventional LLM-based approaches remain poorly aligned with the demands of NAMD analysis. \cite{li2025chemvlmexploringpowermultimodal, chenLargescaleChemicalReaction2025}
Most existing working paradigms apply LLMs as either autonomous planners or passive executors.
Neither role is suitable for trajectory interpretation, which requires simultaneous visual comprehension and chemical reasoning of complex spatial patterns.
Critically, text-only LLMs lack access to the visual cues essential for understanding molecular geometries and reaction pathways, leaving this domain largely unexplored.

To address these challenges, we present VisU, a visual-LLM-driven framework designed to automate the unsupervised analysis of NAMD trajectories.
VisU adopts a ``\textit{Mentor-Engineer-Student}'' paradigm that mimics the collaborative workflow of a theoretical chemistry research group.
This framework integrates two recently-released state-of-the-art large language models,
\textit{i.e.} Doubao-Seed-1.6-Vision (released September 30, 2025) \cite{bytedance2025doubaoseed} and DeepSeek-V3.2 (released November 1, 2025) \cite{deepseekai2025deepseekv32pushingfrontieropen},  to provide a synergistic approach to data analysis.
Within this framework, an experienced \textit{Mentor} provides physical intuition through visual reasoning and domain knowledge.
A programming \textit{Engineer} handles adaptive tool-calling and workflow orchestration, while a \textit{Student} executes computational tasks, generates visualizations, and submits results for evaluation.
This design establishes a research paradigm that integrates multimodal reasoning, visual inspection, and chemical expertise within a fully automated, four-stage workflow for analyzing the NAMD simulation data.

By orchestrating vision-capable LLMs within this structured framework, VisU implements an end-to-end perception-to-insight pipeline that bridges the gap between raw numerical data and interpretable chemical knowledge.
In this framework, visual reasoning is not merely a supplementary tool but the primary driver of the iterative process, mimicking how a scientist's eye discerns complex patterns to refine hypotheses and validate mechanisms.
This work demonstrates that embedding autonomous visual intelligence into the NAMD analysis workflow redefines how we process excited-state dynamics.
At the same time, this novel research paradigm offers a scalable template for human-AI collaboration across theoretical chemistry,
which enables researchers to uncover molecular mechanisms from massive datasets with unprecedented clarity and efficiency.

\section{Results}

\subsection{A Brief Description of VisU}

In this section, we provide an overview of VisU, an autonomous vision-AI-powered unsupervised analysis framework.

In a traditional theoretical team, three roles perform distinct but complementary functions in a joint project:
an experienced mentor supervises the overall analysis and provides expert chemical guidance, a skilled programming-engineer handles computational tool construction and processing script orchestration, and a diligent student executes computational tasks while generating intermediate results.

VisU framework adopts a ``\textit{Mentor-Engineer-Student}'' architecture, in which these three roles are designed faithfully mirroring the division of labor in a real nonadiabatic dynamics research group.
\begin{itemize}
    \item \textbf{The \textit{Mentor}:} Powered by the Doubao-Seed-1.6-Vision LLM, the \textit{Mentor} integrates deep chemical insights with advanced vision-reasoning capabilities to oversee the analytical direction.
    \item \textbf{The \textit{Engineer}:} Built upon the DeepSeek-V3.2 LLM equipped with specialized computational tools and coding templates, the \textit{Engineer} demonstrates robust programming and debugging proficiency.
    \item \textbf{The \textit{Student}:} Responsible for data management, script execution, and visualization, the \textit{Student} serves as the operational hub, providing feedback to the \textit{Mentor} and \textit{Engineer}.
\end{itemize}
Following this human-mimetic paradigm, VisU operationalizes NAMD analysis into a fully automated, four-stage end-to-end protocol:
\begin{itemize}
    \item \textbf{Preprocessing.}
    This stage requires collaboration of the programming \textit{Engineer} and the meticulous \textit{Student}.
    The \textit{Engineer} is responsible for the script development and code debugging, while the \textit{Student} primarily executes them to extract relevant structures from simulation outputs.
    \item \textbf{Recursive channel discovery.}
    This phase focuses on identifying nonadiabatic channels through recursive unsupervised ML analysis.
    The \textit{Student} operates the analysis pipeline, generates plots, and submits the results for the \textit{Mentor}'s review in the automatic manner.
    The \textit{Mentor} then evaluates these plots using visual reasoning to identify the underlying nonadiabatic events, subsequently providing precise guidance on ML algorithm selection and hyperparameter tuning.
    \item \textbf{Analysis of dominant molecular motions.}
    Once the channels are identified, the \textit{Mentor} and \textit{Student} work together to clarify the dominant molecular motions for each decay pathway.
    The \textit{Student} performs specialized programs for downstream analysis.
    The \textit{Mentor} interprets the key molecular motions by examining the dimensionality reduction results provided by the \textit{Student}, and understands the machanisms of these findings based on chemical insights.
    \item \textbf{Final validation and mechanistic summarization.}
    At the end, VisU delivers a highly professional academic presentation under the joint work between the \textit{Student} and the \textit{Mentor}.
    The \textit{Student} organizes and submits all comprehensive findings, including publication-quality visualizations, to the \textit{Mentor}.
    The \textit{Mentor} conducts the final validation of mechanistic consistency and synthesizes the submitted materials into the final structured report and interpretive representation.
\end{itemize}

By integrating multimodal reasoning, visual understanding, and domain-specific knowledge, VisU mirrors the heuristic intelligence of a research team.
It establishes a fully automated end-to-end protocol that uncovers nonadiabatic channels and their underlying mechanisms without manual intervention.
The overall framework is illustrated in Scheme~\ref{fig:fig1}, and each component is described in detail in the following subsections.

To demonstrate its potential, the current implementation of VisU builds upon the validated framework of previous work. \cite{zhuPrincipalComponentAnalysis2022a}
That work established a hierarchical analysis protocol for characterizing ring deformations in atomic systems via a combination of principal component analysis (PCA) \cite{wold1987principal, abdi2010principal} and various clustering algorithms\cite{ester1996density,schubert2017dbscan,kriegel2011density,ward1963hierarchical,kaufman2009finding}.
Future versions will expand the framework's applicability to a broader spectrum of molecular architectures and complex photophysical processes.

\begin{scheme}[H]
    \centering
    \includegraphics[width=1\linewidth]{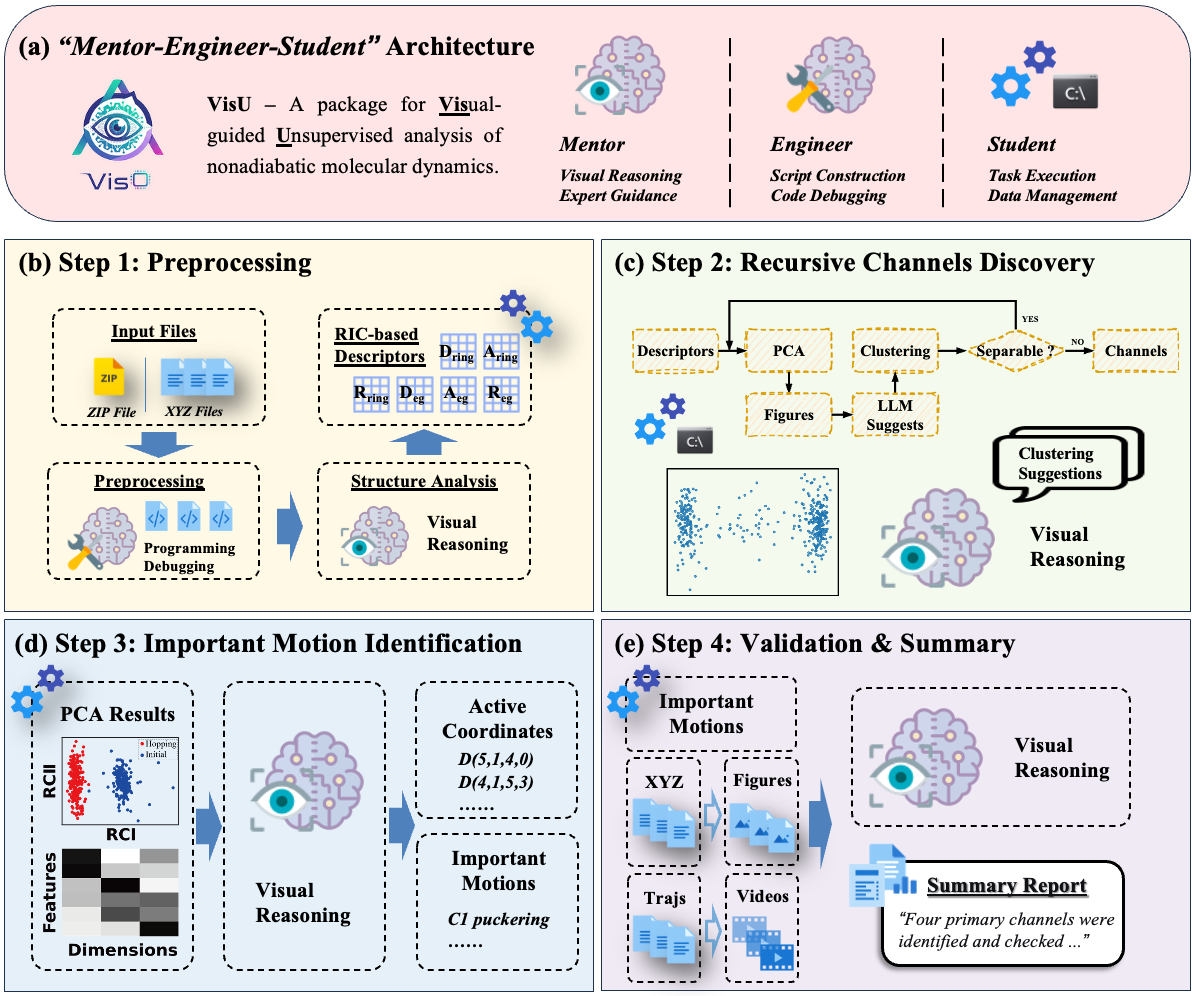}
    \caption{
Workflow of VisU, an autonomous vision-AI framework for hierarchical analysis of nonadiabatic dynamics.
The VisU system comprises the three-role paradigm and four functional modules:
(a) ``\textit{Mentor-Engineer-Student}'' architecture;
(b) \textit{Preprocessing} module autonomously parses raw NAMD outputs and generates standardized descriptor sets;
(c) \textit{Recursive Channel Discovery} module iteratively applies PCA-guided clustering to uncover chemically distinct decay channels;
(d) \textit{Important-Motion Identification} module analyzes PCA projections and visualizes dominant nuclear displacements for each channel;
(e) \textit{Validation and Mechanistic Summary} module selects representative structures, performs final quality checks, and generates concise mechanistic interpretations.
    }
    \label{fig:fig1}
\end{scheme}

\subsubsection{Data Preprocessing}

In the first step of VisU, the \textit{Data Preprocessing} module performs two main tasks.
It generates geometry and other relevant files in a standardized format for subsequent analysis, and converts the geometric information into molecular descriptors.
This stage involves close coordination among the \textit{Mentor}, \textit{Engineer}, and \textit{Student}, combining chemical insight, reliable programming ability, and automated execution.

\subsubsubsection{Input files processing.}
VisU accepts two input formats: (i) three mandatory XYZ files, that is, \texttt{ref.xyz} (reference geometry of ground-state minimum), \texttt{hopping.xyz} (all hopping geometries from NAMD trajectories), and \texttt{init.xyz} (corresponding initial geometries), or (ii) a single ZIP archive containing raw simulation outputs from any NAMD package.
Although both approaches are allowed, the direct submission of the three XYZ files is highly recommended for routine analyses, as it saves tokens and eliminates any remaining risk of extraction errors.

When a ZIP archive containing raw NAMD outputs is provided, the \textit{Data Preprocessing} module extracts essential structural information and organizes it into the three standardized geometry files, \texttt{ref.xyz}, \texttt{hopping.xyz}, and \texttt{init.xyz}.
This task is nontrivial because output files produced by different NAMD packages are often heterogeneous in format, inconsistent in structure, and highly package-specific, posing a significant challenge for defining a unified preprocessing pipeline.
To address this issue, VisU relies on close coordination between the \textit{Engineer} and the \textit{Student}, where the former develops all preprocessing scripts and the latter executes them and provides feedback through automated validation.

In this workflow, a carefully prompted programming \textit{Engineer}, built upon DeepSeek-V3.2, is responsible for generating all preprocessing scripts.
Regardless of the NAMD package used to produce the raw outputs, the \textit{Engineer} dynamically constructs appropriate scripts based solely on the information contained in the ZIP archive.
With minimal user-provided hints, the \textit{Engineer} and \textit{Student} collaboratively extract the relevant data and generate the three standardized geometry files without human supervision.

In general, different NAMD packages produce distinct output formats and directory structures.
Conventional LLM-based workflows lack direct access to file contents and system details, preventing them from reliably interpreting such heterogeneous file systems during preprocessing.
Without knowledge of the file contents or structures, these pipelines typically rely on multistep trial-and-error workflows, iteratively attempting to interpret outputs, generate scripts, execute them, and refine the logic based on error messages.
This process not only incurs high computational costs and carries a substantial risk of non-convergence but also requires extensive elaborate instruction frameworks to handle diverse scenarios.

To overcome these limitations, VisU adopts a skilled \textit{Engineer} based on the state-of-the-art ``Reasoning-in-Action'' paradigm implemented in DeepSeek-V3.2.
By actively deciding which computational tools to invoke during reasoning, the \textit{Engineer} constructs a coherent, global understanding of the output file system while generating preprocessing scripts.
This design significantly improves coding efficiency and reduces the reliance on exhaustive manual prompting.
In practice, a set of well-defined validation tools implemented as Python functions is provided, enabling the \textit{Engineer} to inspect directory structures, read files, and verify generated outputs in a highly automated and efficient manner.

To ensure broad compatibility across the heterogeneous and often inconsistent output file formats generated by different NAMD simulation packages, we deliberately avoid allowing LLMs to generate a complete preprocessing script from scratch.
Instead, the framework provides a curated library of expert-validated Python templates for structure extraction, format sanitization, and data integrity checks.
With the support of these templates, the \textit{Engineer} constructs the full preprocessing script by assembling and adapting predefined components rather than writing arbitrary code.
This design substantially improves both flexibility and reliability in data preprocessing, surpassing conventional LangChain-style agents that are often constrained by static tool graphs or rigid, hard-coded workflows.
Following script generation, the \textit{Student} executes the preprocessing codes, generates all required files, and returns execution feedback.
If execution fails, the complete traceback information is automatically fed back to the LLM, enabling iterative self-repair and refinement of the preprocessing operations until valid output files are produced.
Overall, this template-grounded and self-repairing strategy establishes a more effective paradigm for automated scientific code generation and robust preprocessing of heterogeneous simulation outputs.
Crucially, the tools and templates developed for the \textit{Engineer} are designed to be general-purpose and system-agnostic, rather than being tailored to specific case studies.
More details are provided in the SI.

\subsubsubsection{Structure analysis and descriptor generation.}
Following preprocessing, VisU proceeds to systematic structure analysis and descriptor generation through the close cooperation of the \textit{Mentor} (built on Doubou-Seed-1.6-Vision) and \textit{Student}.
The core of this stage is the introduction of a dedicated vision-language chemistry \textit{Mentor}, a large multimodal model extensively prompted with expert-level chemical knowledge.
The workflow starts with the \textit{Student} generating a high-quality 2D molecular diagram from the optimized reference geometry in \texttt{ref.xyz}.

After an initial validation of the reference structure, the \textit{Mentor} analyzes the molecular building blocks and automatically partitions atoms into chemically meaningful groups, such as ring and non-ring atoms.
Leveraging its multimodal understanding of bonding topology and visual patterns, it identifies and enumerates atoms belonging to aromatic, alicyclic, and fused ring systems.
This fully automated, expert-level structural annotation eliminates manual intervention while ensuring the generation of chemically interpretable descriptors.

As the currently supported analysis workflow in VisU primarily targets the geometric evolution of atomic ring systems in NAMD simulations, the following discussion focuses on such cases.
This scope is sufficient to illustrate the effectiveness of the vision-powered ``\textit{Mentor-Engineer-Student}'' cooperative strategy for NAMD analysis, while extensions to more general molecular topologies are left for future work.

Previous works~\cite{zhuPrincipalComponentAnalysis2022a,liu2025automatedframeworkanalyzingstructural} demonstrated that redundant internal coordinates (RICs) \cite{pulay1992geometry,peng1996using} provide an appropriate molecular descriptor for NAMD analysis and are therefore adopted here.
RICs consist of bond lengths (R), bond angles (A), and dihedral angles (D), which span different numerical ranges and encode distinct physical aspects of molecular structure.
For systems containing atomic rings, RICs associated with the ring framework and those involving terminal groups play fundamentally different roles during nonadiabatic dynamics.
Accordingly, within the workflow, the \textit{Mentor} instructs the \textit{Student} to classify atoms into ring (\textit{ring}) and end-group (\textit{eg}) subsets.
The \textit{Student} then constructs six descriptor groups, $R_{\mathrm{ring}}$, $A_{\mathrm{ring}}$, $D_{\mathrm{ring}}$, $R_{\mathrm{eg}}$, $A_{\mathrm{eg}}$, and $D_{\mathrm{eg}}$, which are subsequently used for unsupervised analysis.
To avoid overemphasizing large-amplitude hydrogen motions, hydrogen atoms are excluded, and the analysis focuses exclusively on the molecular scaffold.

\subsubsection{Recursive Channel Discovery}

The primary objective of this stage is to examine the distribution patterns of hopping geometries and identify the distinct reaction channels within the NAMD trajectories.
This is achieved through a combination of dimensionality reduction and clustering analysis, executed via a collaborative \textit{Mentor-Student} workflow.
Leveraging the multimodal capabilities of the vision-LLM, the \textit{Mentor} acts as an indefatigable chemical data analyst, providing expert guidance to the \textit{Student}, who serves as the primary executor.

The ``PCA-then-clustering'' protocol follows a structured iterative loop:
\begin{itemize}
    \item \textbf{Dimensionality reduction.} The \textit{Mentor} instructs the \textit{Student} to perform PCA on all hopping geometries using six distinct descriptor families and to present the resulting 2D projection figures for validation.
    \item \textbf{Visual selection.} The \textit{Mentor} visually inspects these figures to evaluate their chemical significance and identify the low-dimensional representation that exhibits the most discriminative grouping.
    \item \textbf{Clustering execution.} Based on the identified patterns, the \textit{Mentor} selects an appropriate clustering method and specifies its initial parameters. The \textit{Student} then applies the chosen algorithm to partition the hopping geometries.
\end{itemize}

This three-step protocol is applied recursively.
After each clustering event, the \textit{Student} extracts each resulting subgroup as a new dataset and initiates the protocol again.
This cycle continues recursively until a cluster is deemed non-separable or its size falls below a predefined threshold (typically 15\% of the total hopping geometries).
Upon completion, each terminal cluster is identified as a single candidate nonadiabatic channel.
In implementation, this recursive procedure creates a hierarchical tree that enables the systematic enumeration of all reaction pathways, significantly enhancing the modularity and robustness of the automated analysis framework.

The efficiency of this protocol relies on three critical decision-making tasks:
(i) selecting the optimal reduced descriptor set;
(ii) choosing the most suitable clustering algorithm;
and (iii) fine-tuning clustering parameters.
Although these tasks are largely intuitive for experienced researchers, they are extremely difficult to automate using rigid mathematical rules.
By incorporating vision-enabled reasoning, the \textit{Mentor} can directly inspect visual representations of the data and identify salient patterns that would otherwise be inaccessible to purely numerical criteria.
This capability allows the \textit{Mentor} to evaluate projections and clustering outcomes in a manner closely resembling the visual assessments routinely performed by human theoretical chemists.
Consequently, subsequent analytical decisions are guided by visually grounded scientific judgment rather than fixed heuristics.

The selection of an appropriate descriptor subset follows a standardized visual screening procedure.
The \textit{Student} generates six low-dimensional projections corresponding to all descriptor groups and submits them to the \textit{Mentor} for inspection.
Based on the separability, compactness, and overall structure of the data distributions, the \textit{Mentor} identifies the reduced spaces that exhibit the most distinguishable channel patterns and selects the corresponding reduced space for subsequent clustering analysis.

To facilitate the selection of the clustering algorithm, the \textit{Mentor} is provided with a set of expert-defined principles via specialized prompts.
By ``perceiving'' the data distribution patterns and evaluating them against these principles, the \textit{Mentor} determines the optimal strategy among three supported methods: K-Means \cite{lloyd1982least,macqueen1967some,ikotunKmeansClusteringAlgorithms2023}, Density-Based Spatial Clustering of Applications with Noise (DBSCAN) \cite{ester1996density,schubert2017dbscan,kriegel2011density}, and agglomerative clustering \cite{ward1963hierarchical,kaufman2009finding}.
Further details of these algorithms can be found in the SI.

A particularly critical challenge lies in the dynamic adjustment of clustering parameters.
While humans can instinctively evaluate clustering quality through visual inspection, formalizing this intuition into a unified mathematical objective function remains non-trivial.
VisU overcomes this obstacle by leveraging vision-reasoning to address two distinct operational scenarios.

For algorithms such as K-Means or agglomerative clustering, parameter selection (e.g., the number of clusters, $k$) can be derived almost exclusively from the visual distribution of the data. In these cases, the \textit{Student} performs an initial execution using randomized or default parameters to produce a preliminary labeled scatter plot.
The \textit{Mentor}  then inspects this visualization and directly dictates the necessary adjustments to align the results with the perceived data distribution.

For algorithms like DBSCAN, a more complex challenge arises: its core parameters (\texttt{eps} and \texttt{min\_samples}) lack a direct visual interpretation. Even for human experts, determining optimal values through a single visual inspection is nearly impossible. To address this, VisU implements a reasoning-driven strategy that combines expert guidance with iterative evaluation. The \textit{Mentor} is first informed of the meaning of each parameter and general adjustment principles. Then, by observing the sensitivity of clustering outcomes to parameter variations, the \textit{Mentor} and \textit{Student} engage in a multi-turn interaction, systematically refining parameter values based on visual feedback and logical reasoning until a satisfactory partition of reaction channels is obtained.

\subsubsection{Important-Motion Identification}

Following the identification of all involved nonadiabatic reaction channels, the next task is to elucidate the active molecular motions that drive the system from the Franck-Condon region to the conical intersection.
To this end, the hopping geometries belonging to a specific channel and their corresponding initial structures are analyzed collectively using PCA across various descriptor sets.
By examining whether initial and hopping geometries exhibit clear spatial separation in the PCA projections, the descriptor sets primarily responsible for the nonadiabatic decay can be clarified. Subsequent analysis of the contribution of individual descriptor elements to the principal components allows for the identification of the active coordinates governing the dynamics.
This process relies heavily on the interpretation of PCA results, a task that is intuitive for human researchers but difficult to formalize with conventional computational tools.

The identification of these critical motions is performed through a tight coordination between the \textit{Mentor} and the \textit{Student}.
For each identified channel, the \textit{Student} assembles the initial and hopping geometries, executes PCA, and generates a comprehensive package.
This package includes 2D projection plots, heatmaps of the principal components, and three-view (front, side, and top) renderings of both the reference and averaged hopping structures.
Under the guidance of its vision-reasoning capabilities, the \textit{Mentor} evaluates these multi-faceted figures to pinpoint the active coordinates and leading molecular motions characteristic of the channel.
This \textit{Mentor-Student} joint procedure effectively bridges the gap between unsupervised data clustering and human-interpretable chemical mechanisms, establishing a fully automated NAMD analysis protocol that eliminates the requirement for manual validation.

\subsubsection{Validation and Mechanistic Summary}

In the final stage, VisU consolidates all analytical findings into a comprehensive analysis report.
This comprehensive synthesis is achieved through the integration of the \textit{Mentor}'s visual reasoning and chemical expertise, the \textit{Engineer}'s specialized programming capabilities, and the \textit{Student}'s disciplined execution.

For each reaction channel, the \textit{Student} randomly samples a representative set of hopping geometries (typically ten) and prepares their three-view (front, side, and top) renderings for the \textit{Mentor}.
Upon visual evaluation, the \textit{Mentor} identifies the most characteristic geometry based on previously established analysis results.
Simultaneously, the \textit{Mentor} inspects representative geometries across all channels to identify and merge mirror-symmetric pairs, ensuring a concise and physically sound classification.
Following the selection of representative geometries, the \textit{Engineer} and \textit{Student} collaborate to generate trajectory evolution animations.
The \textit{Engineer} adapts and refines predefined code templates to construct robust visualization scripts, which are then executed by the \textit{Student} to produce high-fidelity videos for the visualizations of typical trajectory evolution.

Finally, all generated assets, including analysis recordings, annotated figures, and optional video animations, are forwarded to the \textit{Mentor}.
At this juncture, the \textit{Mentor} leverages both its visual perception and deep chemical insights to perform a systematic evaluation.
Through rigorous domain-specific prompt design, the \textit{Mentor} is equipped with the specialized knowledge of a senior computational chemist.
This allows the \textit{Mentor} to re-examine the entire workflow: identifying reaction channels, tracing dominant atomic motions, correlating them with active coordinates, and performing cross-channel comparisons.

The multimodal assessment capability of the \textit{Mentor} facilitates the generation of a concise, human-readable summary for each channel.
These summaries, paired with key visualizations, are compiled into a definitive mechanistic report.
This final stage elevates VisU from a mere unsupervised analysis engine to an autonomous computational collaborator that not only extracts essential features from NAMD simulations but also delivers expert-level chemical insights.

Overall, by integrating vision-capable LLMs within a systematically orchestrated prompt framework, VisU performs an end-to-end analysis of NAMD simulation data and produces finalized scientific reports with complete autonomy.
This demonstration proves that vision-driven multimodal intelligence provides a transformative solution for complex scientific problems that traditionally require both human visual intuition and domain-specific expertise.

\subsection{Case study}

In this work, we employed keto isocytosine as a prototype system to illustrate the proposed VisU framework.
As a tautomer of cytosine, one of the canonical DNA bases, the photoinduced dynamics of keto isocytosine was extensively investigated both experimentally and theoretically. \cite{gorb1999theoretical, ha1996quantum, kwiatkowski1997density, vranken1994infrared, shukla2000investigations, bakalska2012comparative, szabla2016ultrafast, hu2017nonadiabatic, segarra2021modelling}
The molecular structure of keto isocytosine with atom labeling is shown in Figure~\ref{fig:preprocess}.(d).
The NAMD simulation data used in this study were taken from our previous work, \cite{zhuPrincipalComponentAnalysis2022a} and relevant details are given in SI as well.

\subsubsection{Preprocess the Input File}
\begin{figure}[H]
    \centering
    \includegraphics[width=0.9\linewidth]{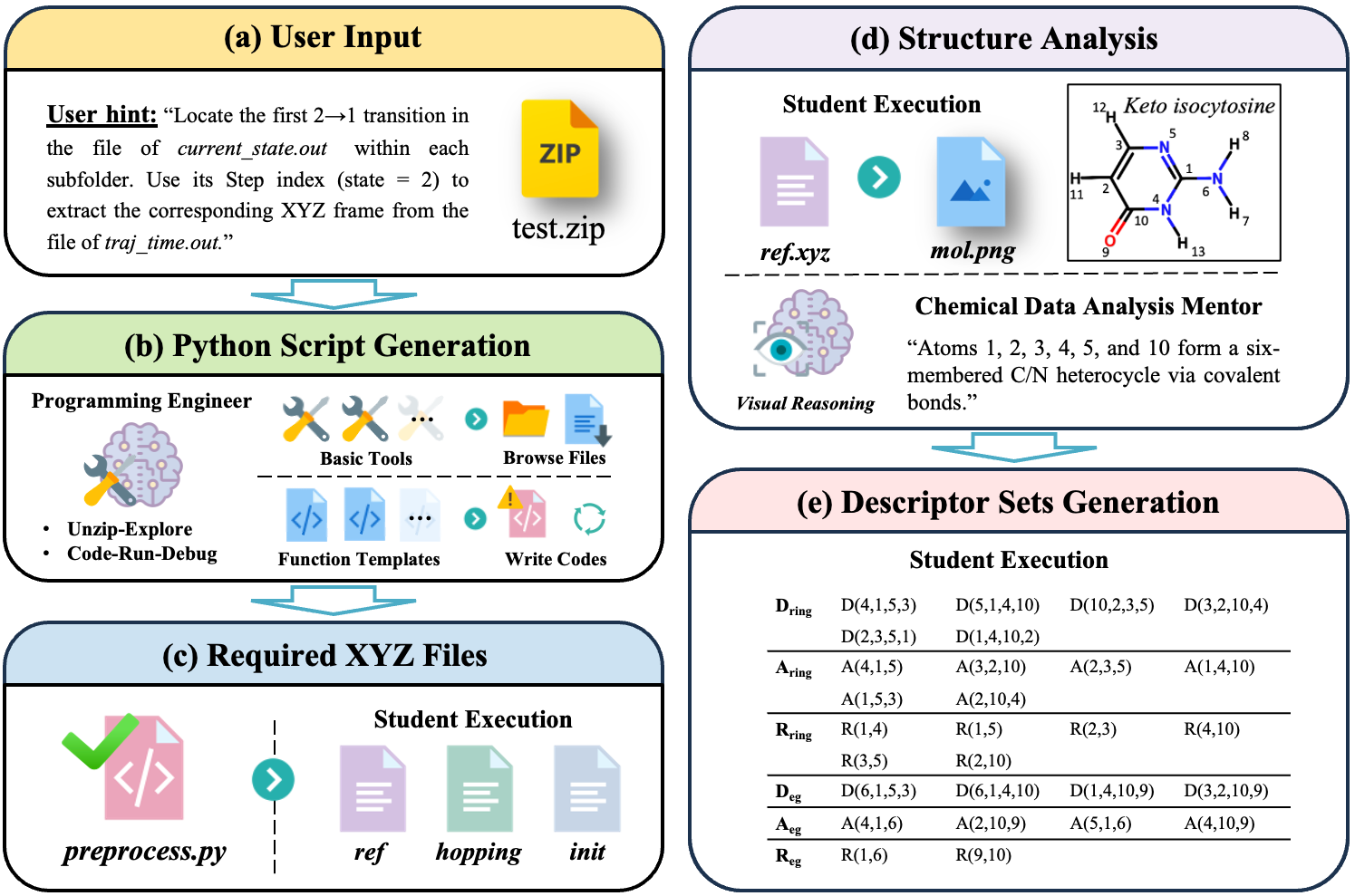}
    \caption{
Preprocessing pipeline implemented through the cooperation of \textit{Mentor}, \textit{Engineer} and \textit{Student}.
(a) User-provided input file (\texttt{test.zip}) and message.
(b) Generation of the preprocessing Python script using tool calling and the provided Python function templates.
(c) Final \texttt{preprocess.py} and the resulting XYZ files.
(d) Analysis of molecular structures, including the molecular structure and atomic labels of keto isocytosine shown in \texttt{mol.png}.
(e) Six generated descriptor sets based on RICs.
    }
    \label{fig:preprocess}
\end{figure}

\subsubsubsection{Preprocess the input ZIP file.}
To demonstrate the end-to-end autonomy of VisU, we deliberately chose the raw, unprocessed NAMD simulation output of from JADE package.
In JADE, each trajectory is stored in a separate subdirectory containing \texttt{current\_state.out} (the current electronic state \textit{v.s.} time) and \texttt{traj\_time.out} (the complete XYZ trajectory from $t = 0$).
For this illustration, the entire simulation workspace was compressed into a single archive, \texttt{test.zip}.

Clearly, as with output files generated by other NAMD packages, the raw JADE data requires extensive preprocessing before formal analysis.
This preprocessing includes unzipping all output files, validating their contents, collecting the initial geometries, identifying the hopping times, and extracting the corresponding hopping geometries.
Here, we emphasize that although the workflow and performance of VisU are demonstrated using keto isocytosine based on JADE outputs, the overall data-processing pipeline of VisU is, in principle, directly applicable to outputs from other NAMD codes without further modification.

After the introduction of the DeepSeek-V3.2, VisU is able to provide a self-correcting and fully automated data-preprocessing pipeline.
As shown in Figure~\ref{fig:preprocess}.(a), the user provides nothing more than the compressed archive and a very brief natural-language hint.
VisU immediately activates the \textit{Engineer} (Figure~\ref{fig:preprocess}.(b)), who dynamically invokes computational tools during reasoning and generates task-specific code.

Rather than directly generating a complete preprocessing script in a single step, the \textit{Engineer} first reasoned about the information required for subsequent analysis, including the directory structure, file naming conventions, and the locations of trajectory and geometry files.
Guided by this reasoning, it selectively invoked file-system inspection tools embedded within VisU to explore the unzipped archive and extract the necessary structural information.
After acquiring the workspace context, the \textit{Engineer} composed a preprocessing script from a curated library of expert-validated function templates.
In this sense, the overall preprocessing procedure closely resembles human expert operations.
The \textit{Student} then executed the script and returned error messages in cases of failure.
In the present example, the program encountered a Python error (\texttt{FileNotFoundError}) during the first execution.
The \textit{Student} collected all error information and relayed it to the \textit{Engineer}, which subsequently revised the script.
The data-preprocessing task was completed successfully after a second iteration.
Details of the error analysis and the corrected preprocessing script are provided in the SI.
As discussed in the previous section, this preprocessing step produced three required output files, namely \texttt{ref.xyz}, \texttt{init.xyz}, and \texttt{hopping.xyz} (Figure~\ref{fig:preprocess}.(c)).

With only minimal input hints, VisU demonstrates strong data-preprocessing capability by automatically handling a previously unseen NAMD output archive.
This successful application can be attributed to two essential features of VisU: the tool-aware reasoning capability enabled by DeepSeek-V3.2 and the well-defined reference templates provided by the developers.

\subsubsubsection{Generate the descriptor set.}
To generate molecular descriptors, VisU processed both the reference (\texttt{ref.xyz}) and hopping (\texttt{hopping.xyz}) geometries.
The \textit{Student} plotted a 2D molecular image (\texttt{mol.png}) with atomic labels in the basis of the reference geometry, as illustrated in Figure~\ref{fig:preprocess}.(d).
This image was then passed to the \textit{Mentor}, powered by the Doubao-Seed-1.6-Vision model, which was carefully prompted to understand the context of the analysis workflow.

Leveraging its visual-reasoning capability, this \textit{Mentor} first interpreted the image to determine whether the input molecule was compatible with the current available protocol.
For the present system, VisU identified the atoms constituting the ring moiety, which provided explicit guidance for the \textit{Student} to construct the corresponding RIC descriptors.
Using the hopping geometries collected in \texttt{hopping.xyz}, the \textit{Student} then built six sets of RIC descriptors, as shown in Figure~\ref{fig:preprocess}.(e): $D_{\mathrm{ring}}$, $A_{\mathrm{ring}}$, $R_{\mathrm{ring}}$, $D_{\mathrm{eg}}$, $A_{\mathrm{eg}}$, and $R_{\mathrm{eg}}$.
These descriptors formed the foundation for subsequent unsupervised channel discovery and mechanistic analysis.
In parallel, the same descriptor sets were computed for all initial geometries in \texttt{init.xyz}.

\subsubsection{Extract the Channels}

\begin{figure}[H]
    \centering
    \includegraphics[width=0.8\linewidth]{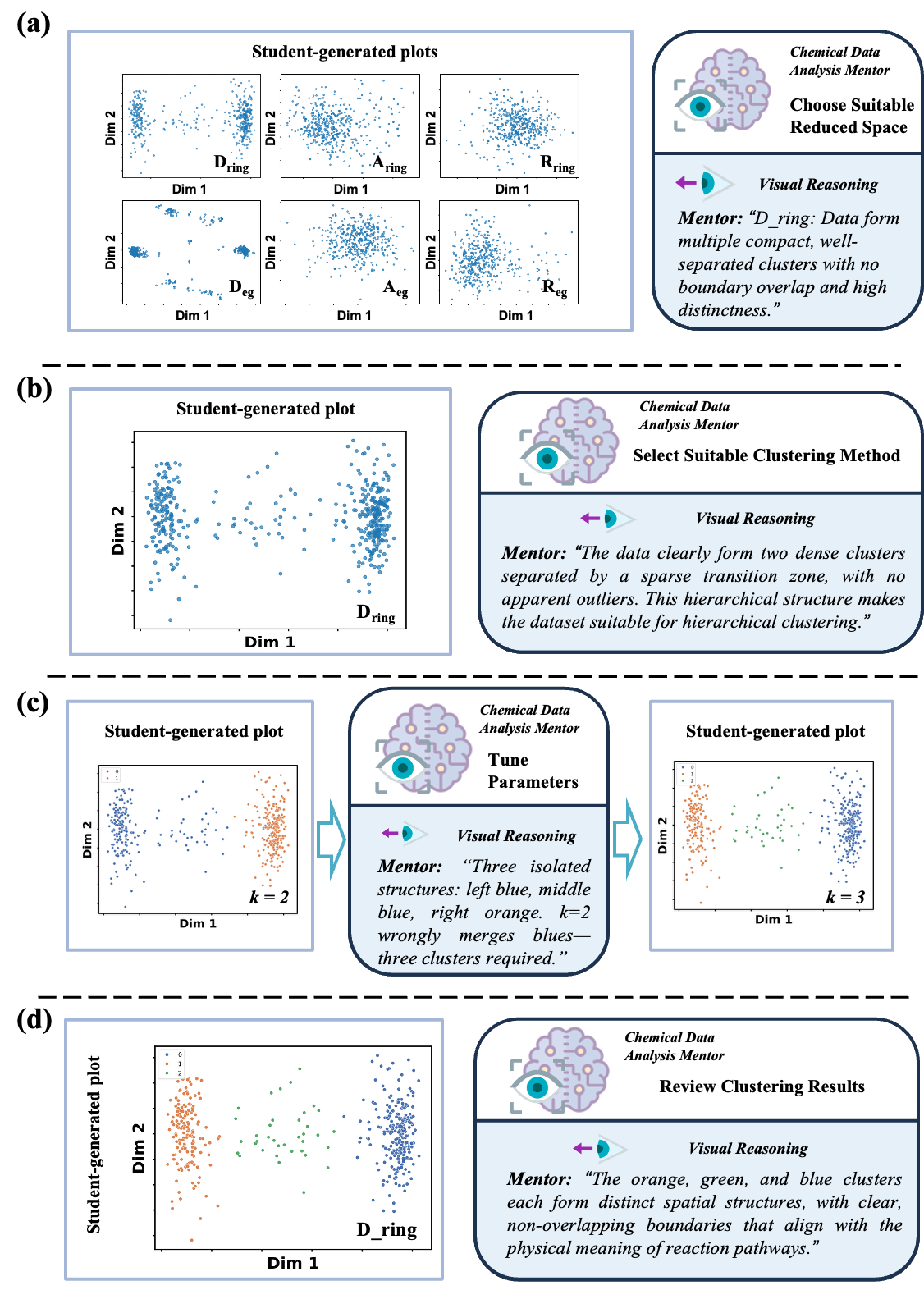}
    \caption{
``\textit{Mentor-Student}'' cooperation during the first iteration of unsupervised channel discovery.
Based on diagnostic plots produced by the \textit{Student}, the carefully prompted visual-language \textit{Mentor} (Doubao-Seed-1.6-Vision) provides targeted guidance on:
(a) selection of the optimal reduced-dimensional subspace,
(b) choice of clustering algorithm,
(c) tuning of hyperparameters,
and (d) final evaluation of the resulting clusters.
    }
    \label{fig:dm_clu_eg}
\end{figure}

Using the prepared descriptor sets, channel discovery proceeded via an iterative PCA-then-clustering analysis protocol \cite{zhuPrincipalComponentAnalysis2022a}.

As shown in Figure~\ref{fig:dm_clu_eg}.(a), the first iteration of channel discovery began by applying PCA to each of the six descriptor sets to obtain low-dimensional representations of the hopping geometries.
In all cases, the first two principal components contributed more than 60\% of the total variance.
The resulting PCA projections were then forwarded to the \textit{Mentor} for assessment.
In the present example, the \textit{Mentor} leveraged its visual-reasoning capability to identify that the hopping points were well separated in the $D_{\mathrm{ring}}$-based reduced space, indicating the presence of several distinct channels and highlighting the important role of ring-associated dihedral motions in the nonadiabatic dynamics.

The next critical step is to group similar hopping geometries into clusters.
In Figure~\ref{fig:dm_clu_eg}.(b), the \textit{Mentor} inspected the distribution in the $D_{\mathrm{ring}}$ PCA plot using its visual-reasoning capability and recommended agglomerative clustering for the initial partitioning.
As shown in Figure~\ref{fig:dm_clu_eg}.(c), the \textit{Student} first performed clustering with the default parameter $k = 2$.
Upon reviewing the resulting plot, the \textit{Mentor} immediately identified that the central region was visually distinct and was not be merged with the adjacent cluster.
Therefore, the \textit{Mentor} instructed the \textit{Student} to adjust the parameter to $k = 3$, after which the clustering result aligned with expectations.
Following re-clustering with the corrected parameter, the \textit{Mentor} confirmed that the current clustering was accurate.

A final visual inspection by \textit{Mentor}, shown in Figure~\ref{fig:dm_clu_eg}.(d), verified that the current clustering result with three clusters was reasonable, ensuring that this step aligned with expert-level visual intuition.
The first discovery iteration thus concluded with three clusters: \textit{Cluster 0}, \textit{Cluster 1}, and \textit{Cluster 2}.
Among them, \textit{Cluster 2} contained only 37 structures, corresponding to less than 15\% of the total dataset.
This cluster was therefore treated as an independent candidate channel (\textit{Candidate Channel 0}), while the remaining two clusters proceeded to a second iteration of analysis.
Afterward, the same PCA-then-clustering procedure was then recursively applied to all remaining subsets.
As illustrated in Figure~S1, all clusters either became non-separable after three iterations or contained fewer than 15\% of the structures, ultimately yielding seven nonadiabatic decay channels (\textit{Candidate Channel 0} to \textit{6}).

As different data patterns appeared in these iterative steps, distinct clustering algorithms were recommended based on their distributions: including agglomerative clustering (Iteration 1), DBSCAN (Iteration 2), and K-Means (Iteration 3), as in Figure~S1.
Additional results are provided in the SI.
These observations demonstrate that VisU can effectively handle both algorithms like agglomerative clustering, which requires only a single hyperparameter and aligns closely with visual intuition, and more complex methods such as DBSCAN, where the two parameters (\texttt{eps} and \texttt{min\_samples}) are less directly interpretable from visual inspection.

\subsubsection{Identify the Important Molecular Motion}

\begin{figure}[H]
    \centering
    \includegraphics[width=0.9\linewidth]{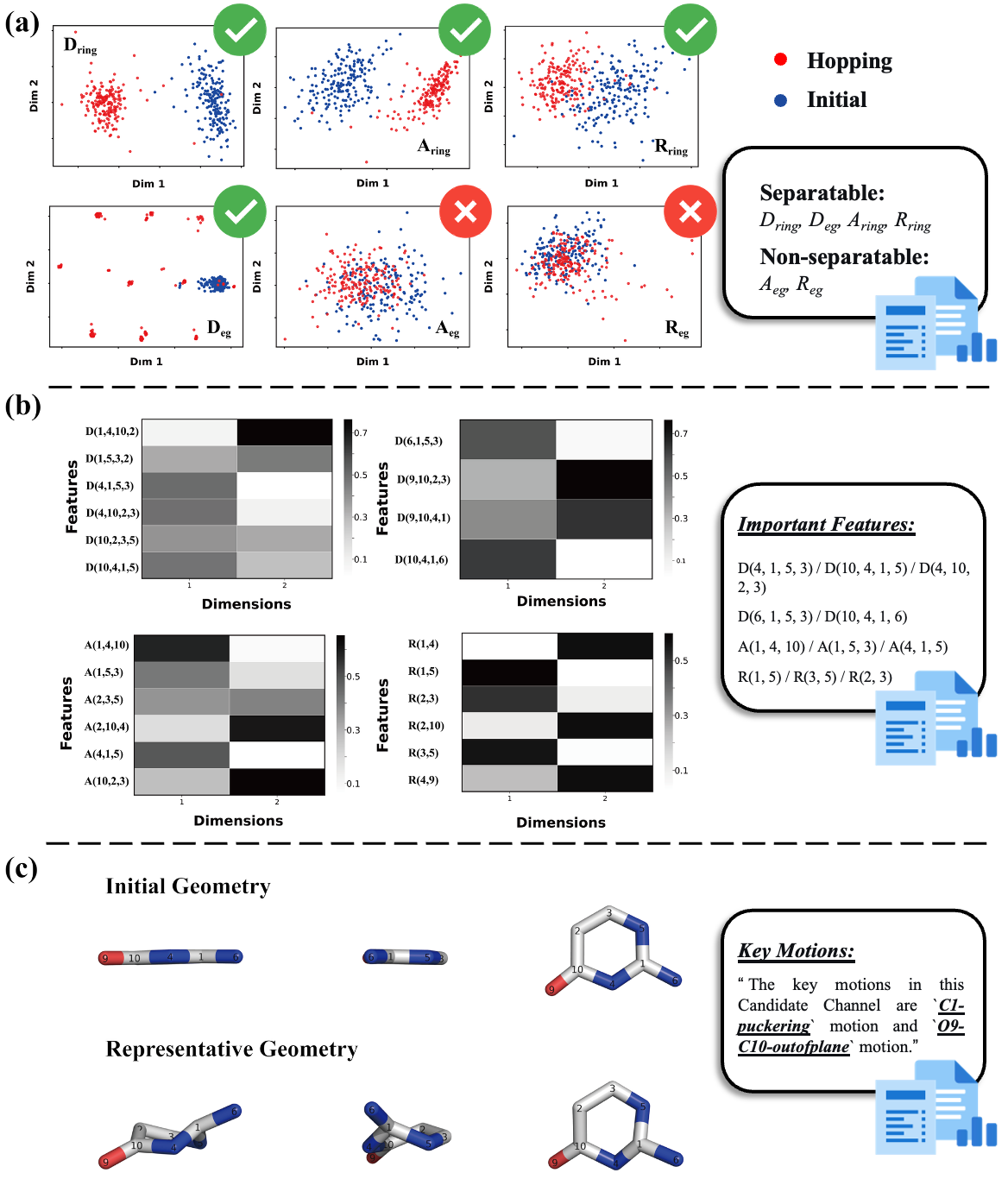}
    \caption{
Identification of the dominant molecular motions, illustrated for \textit{Candidate Channel 3}.
(a) Two-dimensional PCA projections generated by the \textit{Student} from the structural descriptors $D_{\mathrm{ring}}$, $A_{\mathrm{ring}}$, $R_{\mathrm{ring}}$, $D_{\mathrm{eg}}$, $R_{\mathrm{eg}}$, and $A_{\mathrm{eg}}$.
By visually inspecting these projections, the \textit{Mentor} assessed in which low-dimensional representations the initial and hopping geometries became separable.
(b) Corresponding PCA component loadings, examined by the \textit{Mentor} to identify the structural features most responsible for the observed channel separation.
(c) Dominant molecular motions inferred by the \textit{Mentor} through joint inspection of the key features and representative initial and hopping geometries.
    }
    \label{fig:feat}
\end{figure}

After extracting the candidate channels, the important molecular motions associated with each channel were identified.
\textit{Candidate Channel 2} is presented here as a representative example to illustrate the analysis workflow and the resulting mechanistic interpretation (Figure~\ref{fig:feat}).

At this stage, the \textit{Student} performed PCA projections of both the initial and hopping geometries, submitting the resulting two-dimensional representations to the \textit{Mentor} for analysis.
By visually inspecting these projection plots, the \textit{Mentor} determined which sets of RICs were most effective for distinguishing hopping from initial geometries.
In the present example (Figure~\ref{fig:feat}.(a)), the hopping and initial geometries were clearly separated in the reduced space defined by $D_{\mathrm{ring}}$, $D_{\mathrm{eg}}$, $A_{\mathrm{ring}}$, and $R_{\mathrm{ring}}$, indicating that these RIC sets captured the leading reactive coordinates in the photochemical processes.

Subsequently, the \textit{Student} computed the contribution of each RIC to the principal components and visualized the results in Figure~\ref{fig:feat}.(b).
Upon reviewing these plots, the \textit{Mentor} identified the most significant reactive coordinates, including D(4,1,5,3), D(10,4,1,5), and D(4,10,2,3) in $D_{\mathrm{ring}}$; D(6,1,5,3) and D(10,4,1,6) in $D_{\mathrm{eg}}$; A(1,4,10), A(1,5,3), and A(4,1,5) in $A_{\mathrm{ring}}$; as well as R(1,5), R(3,5), and R(2,3) in $R_{\mathrm{ring}}$.

As shown in Figure~\ref{fig:feat}.(c), using these key features, the \textit{Mentor} further identified that the dominant molecular motions governing \textit{Candidate Channel 2} are \textit{C1 puckering} and \textit{O9-C10 out-of-plane} deformation.
With these identified critical motions, the \textit{Mentor} then selected a representative geometry for mechanistic inspection from several randomly chosen hopping snapshots collected in the previous steps.

After the \textit{Mentor} selected representative hopping geometries, the \textit{Engineer} generated a Python script to extract the corresponding trajectory structures and produce videos for visual inspection of the molecular evolution in each channel.

The same analysis procedure was systematically applied to all candidate channels, enabling the identification of their respective dominant molecular motions, representative structures, and corresponding trajectory animations.
All additional details are provided in the SI.

\subsubsection{Validate and Summarize Mechanisms}

\begin{figure}[H]
    \centering
    \includegraphics[width=0.9\linewidth]{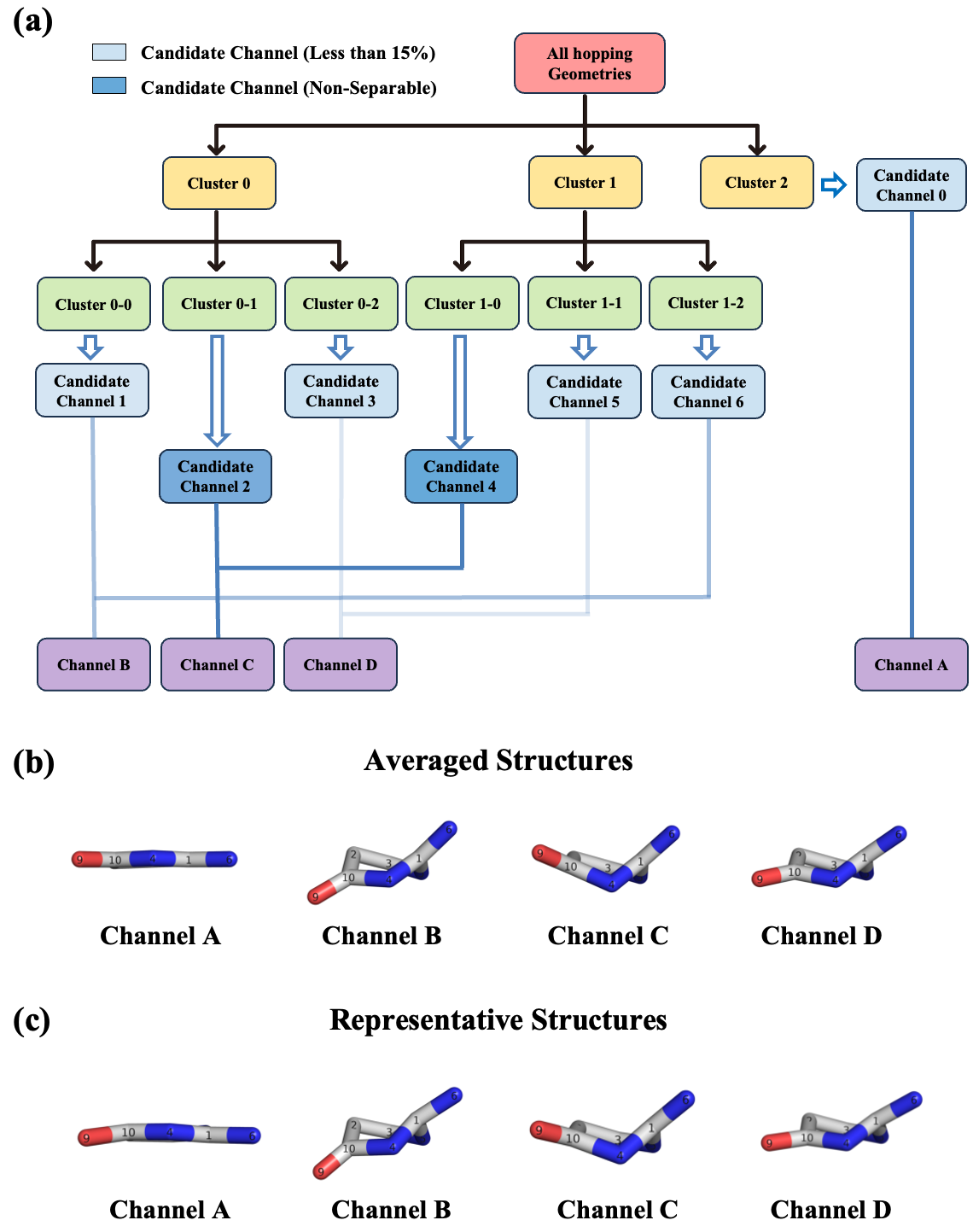}
    \caption{
(a) Final nonadiabatic channel analysis tree. Details regarding the division and naming of different clusters can be found in the SI.
(b) Averaged and (c) representative structures (hydrogen atoms omitted) for the four channels.
    }
    \label{fig:all_results}
\end{figure}

Up to this point, seven candidate nonadiabatic channels and their characteristic molecular motions were identified.
The next step involves the \textit{Mentor} evaluating the results to assess the rationality of all discoveries and to prepare the final analysis report.

An essential task for the \textit{Mentor} is to check for chirality.
For example, several candidate channels in the present system were identified as mirror counterparts based on both qualitative visual inspection and quantitative data comparison.
As shown in Figure~\ref{fig:all_results}.(a), the \textit{Mentor} merged different groups showing mirror-symmetry:
\textit{Candidate Channel 1} and \textit{Candidate Channel 6} into \textit{Channel B};
\textit{Candidate Channel 2} and \textit{Candidate Channel 4} into \textit{Channel C};
and \textit{Candidate Channel 3} and \textit{Candidate Channel 5} into \textit{Channel D}.
\textit{Candidate Channel 0} exhibited no mirror-symmetric counterpart and was therefore retained as \textit{Channel A}.
As in Figures~\ref{fig:all_results}.(b) and (c), the initial seven candidate channels were finally consolidated into four distinct nonadiabatic channels: \textit{Channel A}, \textit{B}, \textit{C}, and \textit{D}.

It is important to note that different decay channels may be governed by the same nuclear motions but with varying magnitudes.
For instance, \textit{Channel B}, \textit{C}, and \textit{D} all exhibited pronounced out-of-plane motions of O9 and C10, while their direction differed.
In principle, one can perform systematic pairwise comparisons of initial and hopping geometries across channels; however, this approach becomes tedious in multichannel situations.
Instead, VisU relies on the \textit{Mentor} to perform a holistic comparison by integrating key features, representative geometries, and dominant motions across all channels.
Through direct visual inspection and cross-channel assessment, the \textit{Mentor} clarifies the mechanistic similarities and differences among the nonadiabatic decay channels.

Finally, the \textit{Mentor} drew the following conclusions.
\textit{Channel A} accounted for approximately 9\% of the total hopping events, which was characterized by a dominant \textit{C9-C10 stretching} motion.
In contrast, the remaining three channels involved combinations of \textit{C1 puckering} and \textit{O9-C10 out-of-plane} motions.
Among them, \textit{Channel B}, \textit{Channel C}, and \textit{Channel D} contributed approximately 10\%, 75\%, and 6\% of the total population, respectively.
Although these three channels shared similar core motions, they differed in the directionality of the side-group motion associated with the O9-C10 fragment, as shown in Figures~\ref{fig:all_results}.(b) and (c).
These results are consistent with our previous and recent studies \cite{zhuPrincipalComponentAnalysis2022a,liu2025automatedframeworkanalyzingstructural}, and additional discussion are provided in the SI.

At the end, the complete mechanistic interpretation, including channel populations, representative structures, and visual summaries derived from the raw nonadiabatic dynamics trajectories, was compiled in a comprehensive analysis report provided in the SI.
With this, the end-to-end analysis performed by VisU is complete, where was executed fully automatically without human intervention.

\section{Conclusion}

We developed VisU, a visual-LLM-driven framework for fully automated, unsupervised analysis of nonadiabatic molecular dynamics (NAMD) trajectories.
By emulating the collaborative workflow of a research group through a three-tier ``\textit{Mentor-Engineer-Student}'' cooperation, VisU establishes a new research paradigm that systematically integrates visual reasoning with domain-specific chemical knowledge, enabling the identification of dominant decay channels in the nonadiabatic dynamics and their underlying mechanisms of photoinduced reactions.

The powerful analysis capability of VisU is stems from the integration of two state-of-the-art large language models: Doubao-Seed-1.6-Vision, which provides expert-level visual reasoning, and DeepSeek-V3.2, which enables adaptive tool-calling and script construction.
This synergy facilitates a multiagent paradigm that orchestrates a comprehensive four-module workflow: \textit{Preprocessing}, \textit{Recursive Channel Discovery}, \textit{Important-Motion Identification}, and \textit{Validation and Mechanistic Summary}.

In a representative case study, VisU successfully identified four distinct nonadiabatic channels, uncovered the corresponding leading reactive coordinates, and automatically handled mirror-symmetry merging.
Furthermore, it performed cross-channel comparisons and generated a comprehensive, figure-rich academic report without human intervention.
The framework demonstrated a professional-grade ability to reproduce dominant mechanistic features reported in previous work \cite{zhuPrincipalComponentAnalysis2022a,liu2025automatedframeworkanalyzingstructural}.
Remarkably, starting only from raw NAMD simulation outputs and minimal text prompts, the entire protocol is executed through autonomous visual inspection and expert-like guidance, powered by advanced vision-reasoning LLMs.
These results demonstrate that VisU offers a fully automated, human-like approach for the mechanistic interpretation of NAMD trajectories, substantially lowering the barrier to routine photochemical analysis.

VisU demonstrates that visual reasoning is not merely a supplementary tool but a primary driver for mechanistic discovery in theoretical chemistry.
This framework successfully bridges the long-standing gap between visual experience (the qualitative patterns recognized by experts) and numerical experience (the quantitative data from theoretical simulations).
By integrating multimodal intelligence, VisU transcends traditional data processing and more closely emulates the working patterns of a team of computational researchers, providing professional academic analysis with the higher-level perception and reasoning required for complex scientific discovery.
This vision-driven framework not only redefines the protocol for analyzing complex data in excited-state dynamics simulations, but also provides a transformative blueprint for the whole AI4Chemistry field.
Ultimately, VisU represents a shift toward an agentic research paradigm by seamlessly combining visual reasoning and domain-specific chemical knowledge, empowering researchers to uncover hidden molecular mechanisms with unprecedented efficiency and clarity in the data-rich era.

\section{Methods}

\subsection{Framework Implementation and Environment}

VisU was implemented in Python 3.12. Core dependencies included Scikit-learn (v1.5.1) \cite{pedregosa2011scikit} for dimensionality reduction and clustering, RDKit (v2025.09.1) \cite{landrum2025rdkit} for molecular manipulation, as well as Matplotlib (v3.10.7) \cite{Matplotlib} for plotting.

Molecular geometries in \texttt{hopping.xyz} were aligned to the reference geometry in \texttt{ref.xyz} using the Kabsch algorithm \cite{kabsch1976solution, kabsch1978discussion}.
Redundant internal coordinates (RICs) were generated using Pysisyphus (v1.0.0) \cite{pysisyphus} to represent molecular motions efficiently.

All molecular structures were rendered with PyMOL (v3.1.0) \cite{pymol}.
Optional trajectory animations were generated as MP4 video using PyMOL (v3.1.0) \cite{pymol} and FFmpeg (v7.1.1) \cite{tomar2006converting}.

\subsection{LLM Agents and Workflow Automation}
The ``\textit{Mentor-Engineer-Student}'' agents were powered by state-of-the-art large language models: Doubao-Seed-1.6-Vision (released September 30, 2025) served as the multimodal \textit{Mentor}, while DeepSeek-V3.2 (released November 1, 2025) served as the programming \textit{Engineer}.
API calls were handled via the Volcengine Python SDK Ark \cite{volcengine2025volcengine} for the former and the OpenAI Python SDK \cite{openai2025openai} for the latter.

Doubao-Seed-1.6-Vision provides a substantial advantage in handling complex multimodal reasoning tasks.
Leveraging its 256k-token context window, the model could accommodate multiple trajectory projections, long textual descriptions, and optional trajectory videos within a single prompt for comprehensive analysis.
Context caching ensured the automatic persistence of historical information, while structured outputs enabled efficient post-processing and downstream integration.
Within VisU, these capabilities allow the \textit{Mentor} to inspect multiple candidate channels concurrently, evaluate structural patterns, and generate precise, human-like mechanistic guidance.

DeepSeek-V3.2 integrates advanced autonomous tool-calling within a chain-of-thought (CoT) reasoning paradigm.
In VisU, it empowers the \textit{Engineer} agent to adaptively generate and execute context-aware preprocessing scripts and analysis codes, dynamically interact with the file system, and coordinate with \textit{Student} modules to ensure correct task execution.
These features enable the automated handling of heterogeneous simulation outputs, the flexible adaptation to previously unseen data archives, and the efficient construction of all computational codes and scripts.

Doubao-Seed-1.6-Vision and DeepSeek-V3.2 are adopted as representative state-of-the-art backends, taking into account both performance and cost efficiency.
Importantly, the VisU framework can readily accommodate updated or alternative models, allowing seamless replacement or upgrading to other advanced models as they become available without modification of the overall workflow.

\section*{Supplementary Information Available}
Additional details are provided, including: methodological descriptions of PCA, clustering algorithms, and the \textit{Engineer} agent design; comprehensive results of channel discovery, important-motion identification, and hyperparameter tuning; the final VisU-delivered mechanistic reports; and the example of VisU-generated codes.

\section{Author contributions}

Yifei Zhu and Jiahui Zhang contributed equally to this work.
Yifei Zhu: Conceptualization, Methodology, Software, Formal analysis, Investigation, Data Curation, Writing - original draft, Writing - review \& editing.
Jiahui Zhang: Conceptualization, Methodology, Software, Formal analysis, Investigation, Data Curation, Writing - original draft, Writing - review \& editing.
Binni Huang: Data Curation, Investigation, Visualization.
Zhenggang Lan: Conceptualization, Supervision, Project administration, Resources, Funding acquisition, Writing - review \& editing.

\section{Conflicts of interest}
The authors declare no competing financial interest.

\section{Data Availability}
The structural data for the keto isocytosine case study, including the hopping geometries and corresponding initial ones extracted from nonadiabatic dynamics, are available on our group's website (\url{https://langroup.site/ketoisocytosine/}).
Additional related data and details on the nonadiabatic dynamics simulation settings can also be found on the same website.

Our source codes are publicly available at the Github repository (\url{https://github.com/Yifei-Zhu/VisU.git}).
To run VisU, users must provide their own API key and service URL by configuring the \texttt{.env} file accordingly.

\begin{acknowledgement}
    The authors express sincerely thanks to the National Natural Science Foundation of China (No. 22333003 and 22361132528) for financial support.
    Some calculations in this paper were done on SunRising-1 computing environment in Supercomputing Center, Computer Network Information Center, CAS.
\end{acknowledgement}

\begin{tocentry}
\includegraphics[width=8.4cm, keepaspectratio]{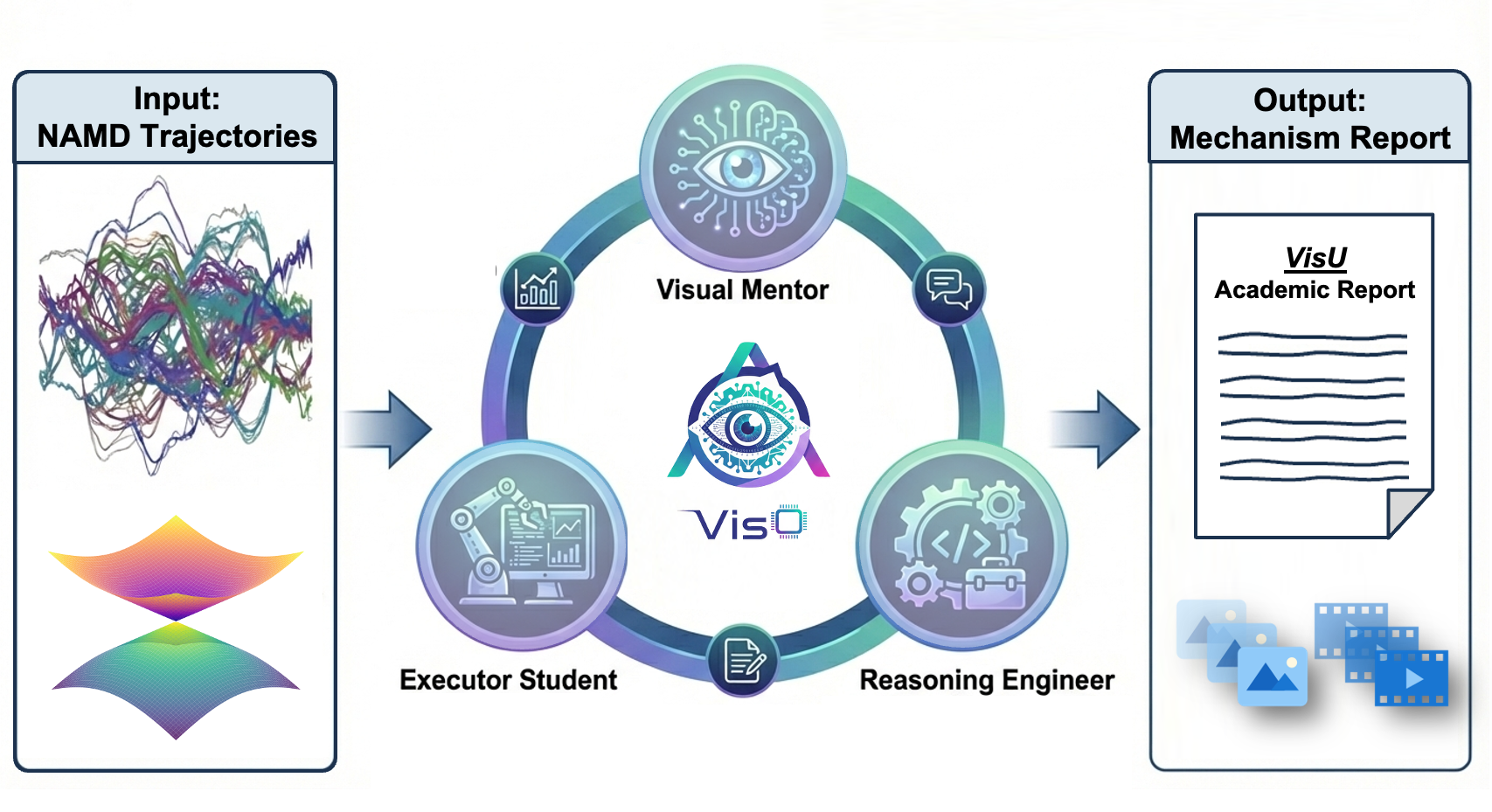}
\end{tocentry}

\bibliography{visu}

@article{pinheiroULaMDynEnhancingExcitedstate2025,
  title = {{{ULaMDyn}}: Enhancing Excited-State Dynamics Analysis through Streamlined Unsupervised Learning},
  shorttitle = {{{ULaMDyn}}},
  author = {Pinheiro, Max and De Oliveira Bispo, Matheus and Mattos, Rafael S. and Telles Do Casal, Mariana and Chandra Garain, Bidhan and Toldo, Josene M. and Mukherjee, Saikat and Barbatti, Mario},
  year = 2025,
  journal = {Digit. Discov.},
  volume = {4},
  number = {3},
  pages = {666--682},
  issn = {2635-098X},
  doi = {10.1039/D4DD00374H}
}

@article{linPredictionExcitedstateReaction2021,
  title = {Prediction of the Excited-State Reaction Channels in Photo-Induced Processes of Nitrofurantoin Using First-Principle Calculations and Dynamics Simulations},
  author = {Lin, Kunni and Hu, Deping and Peng, Jiawei and Xu, Chao and Gu, Feng Long and Lan, Zhenggang},
  year = 2021,
  month = oct,
  journal = {Chemosphere},
  volume = {281},
  pages = {130831},
  issn = {00456535},
  doi = {10.1016/j.chemosphere.2021.130831},
  lccn = {8.9432}
}

@article{hu2017nonadiabatic,
  title={Nonadiabatic dynamics simulation of keto isocytosine: a comparison of dynamical performance of different electronic-structure methods},
  author={Hu, Deping and Liu, Yan Fang and Sobolewski, Andrzej L and Lan, Zhenggang},
  journal={Phys. Chem. Chem. Phys.},
  volume={19},
  number={29},
  pages={19168--19177},
  year={2017},
  publisher={Royal Society of Chemistry}
}

@article{zhuPrincipalComponentAnalysis2022a,
  title = {The Principal Component Analysis of the Ring Deformation in the Nonadiabatic Surface Hopping Dynamics},
  author = {Zhu, Yifei and Peng, Jiawei and Kang, Xu and Xu, Chao and Lan, Zhenggang},
  year = 2022,
  journal = {Phys. Chem. Chem. Phys.},
  volume = {24},
  number = {39},
  pages = {24362--24382},
  issn = {1463-9076, 1463-9084},
  doi = {10.1039/D2CP03323B}
}

@incollection{zhu2022analysis,
  title = {Analysis of Nonadiabatic Molecular Dynamics Trajectories},
  booktitle = {Quantum Chemistry in the Age of Machine Learning},
  author = {Zhu, Yifei and Peng, Jiawei and Liu, Hong and Lan, Zhenggang},
  year = 2022,
  pages = {619--651},
  publisher = {Elsevier}
}

@article{zhuUnsupervisedMachineLearning2024,
  title = {Unsupervised {{Machine Learning}} in the {{Analysis}} of {{Nonadiabatic Molecular Dynamics Simulation}}},
  author = {Zhu, Yifei and Peng, Jiawei and Xu, Chao and Lan, Zhenggang},
  year = 2024,
  month = sep,
  journal = {J. Phys. Chem. Lett.},
  volume = {15},
  number = {38},
  pages = {9601--9619},
  issn = {1948-7185, 1948-7185},
  doi = {10.1021/acs.jpclett.4c01751},
  copyright = {https://doi.org/10.15223/policy-029}
}

@misc{pymol,
  author={Schr\"{o}dinger, LLC and Warren DeLano},
  title={PyMOL},
  url={http://www.pymol.org/pymol},
  version = {2.4.0},
  date = {2020-05-20},
}

@article{liAnalysisTrajectorySimilarity2018,
  title = {Analysis of Trajectory Similarity and Configuration Similarity in On-the-Fly Surface-Hopping Simulation on Multi-Channel Nonadiabatic Photoisomerization Dynamics},
  author = {Li, Xusong and Hu, Deping and Xie, Yu and Lan, Zhenggang},
  year = 2018,
  month = dec,
  journal = {J. Chem. Phys.},
  volume = {149},
  number = {24},
  pages = {244104},
  issn = {0021-9606, 1089-7690},
  doi = {10.1063/1.5048049}
}

@article{gorb1999theoretical,
  title={A theoretical investigation of tautomeric equilibria and proton transfer in isolated and monohydrated cytosine and isocytosine molecules},
  author={Gorb, L and Podolyan, Y and Leszczynski, J},
  journal={J. Mol. Struct.},
  volume={487},
  number={1-2},
  pages={47--55},
  year={1999},
  publisher={Elsevier}
}

@article{ha1996quantum,
  title={Quantum chemical study of structure and stability of all 14 isomers of isocytosine},
  author={Ha, Tae-Kyu and Keller, HJ and Gunde, R and Gunthard, HH},
  journal={J. Mole. Struct.},
  volume={376},
  number={1-3},
  pages={375--397},
  year={1996},
  publisher={Elsevier}
}

@article{kwiatkowski1997density,
  title={Density functional theory study on molecular structure and vibrational IR spectra of isocytosine},
  author={Kwiatkowski, J{\'o}zef S and Leszczynski, Jerzy},
  journal={Int. J. Quantum Chem.},
  volume={61},
  number={3},
  pages={453--465},
  year={1997},
  publisher={Wiley Online Library}
}

@article{vranken1994infrared,
  title={Infrared spectra and tautomerism of isocytosine; an ab initio and matrix isolation study},
  author={Vranken, Hertwig and Smets, Johan and Maes, Guido and Lapinski, Leszek and Nowak, Maciej J and Adamowicz, Ludwik},
  journal={Spectrochim. Acta A},
  volume={50},
  number={5},
  pages={875--889},
  year={1994},
  publisher={Elsevier}
}

@article{shukla2000investigations,
  title={Investigations of the excited-state properties of isocytosine: An ab initio approach},
  author={Shukla, Manoj K and Leszczynski, Jerzy},
  journal={Int. J. Quantum Chem.},
  volume={77},
  number={1},
  pages={240--254},
  year={2000},
  publisher={Wiley Online Library}
}

@article{bakalska2012comparative,
  title={Comparative study of the relaxation mechanisms of the excited states of cytosine and isocytosine},
  author={Bakalska, Rumyana I and Delchev, Vassil B},
  journal={J. Mol. Model.},
  volume={18},
  number={12},
  pages={5133--5146},
  year={2012},
  publisher={Springer}
}

@article{szabla2016ultrafast,
  title={Ultrafast excited-state dynamics of isocytosine},
  author={Szabla, Rafa{\l} and G{\'o}ra, Robert W and {\v{S}}poner, Ji{\v{r}}{\'\i}},
  journal={Phys. Chem. Chem. Phys.},
  volume={18},
  number={30},
  pages={20208--20218},
  year={2016},
  publisher={Royal Society of Chemistry}
}

@article{segarra2021modelling,
  title={Modelling Photoionisation in Isocytosine: Potential Formation of Longer-Lived Excited State Cations in its Keto Form},
  author={Segarra-Mart{\'\i}, Javier and Bearpark, Michael J},
  journal={ChemPhysChem},
  volume={22},
  number={21},
  pages={2172--2181},
  year={2021},
  publisher={Wiley Online Library}
}

@misc{openai2025openai,
  author       = {{OpenAI}},
  title        = {OpenAI Python SDK},
  year         = {2025},
  version      = {2.8.0},
  publisher    = {GitHub},
  url          = {https://github.com/openai/openai-python}
}

@misc{volcengine2025volcengine,
  author       = {VolcEngine/ByteDance-AI},
  title        = {VolcEngine Python SDK},
  year         = {2025},
  note         = {Version 4.0.34},
  howpublished = {GitHub},
  url          = {https://github.com/volcengine/volcengine-python-sdk}
}

@article{pysisyphus,
author = {Steinmetzer, Johannes and Kupfer, Stephan and Gräfe, Stefanie},
title = {pysisyphus: Exploring potential energy surfaces in ground and excited states},
journal = {Int. J. Quantum Chem.},
volume = {121},
number = {3},
pages = {e26390},
doi = {https://doi.org/10.1002/qua.26390},
url = {https://onlinelibrary.wiley.com/doi/abs/10.1002/qua.26390},
eprint = {https://onlinelibrary.wiley.com/doi/pdf/10.1002/qua.26390},
year = {2021}
}

@Article{kabsch1978discussion,
  author    = {Kabsch, Wolfgang},
  title     = {A discussion of the solution for the best rotation to relate two sets of vectors},
  number    = {5},
  pages     = {827--828},
  volume    = {34},
  journal   = {Acta Crystallogr. A Cryst. Phys. Diffr. Theor. Gen. Crystallogr.},
  publisher = {International Union of Crystallography},
  year      = {1978},
}

@Article{kabsch1976solution,
  author    = {Kabsch, Wolfgang},
  title     = {A Solution for the Best Rotation to Relate Two Sets of Vectors},
  number    = {5},
  pages     = {922--923},
  volume    = {32},
  journal   = {Acta Crystallogr. A Cryst. Phys. Diffr. Theor. Gen. Crystallogr.},
  publisher = {International Union of Crystallography},
  year      = {1976},
}

@article{pedregosa2011scikit,
  title={Scikit-learn: Machine learning in Python},
  author={Pedregosa, Fabian and Varoquaux, Ga{\"e}l and Gramfort, Alexandre and Michel, Vincent and Thirion, Bertrand and Grisel, Olivier and Blondel, Mathieu and Prettenhofer, Peter and Weiss, Ron and Dubourg, Vincent and others},
  journal={J. Mach. Learn. Res.},
  volume={12},
  pages={2825--2830},
  year={2011},
  publisher={JMLR. org}
}

@misc{landrum2025rdkit,
  author    = {{Greg Landrum} and {RDKit Contributors}},
  title     = {RDKit: Open-source cheminformatics software},
  year      = {2025},
  version   = {2025.09.1},
  publisher = {GitHub},
  url       = {https://github.com/rdkit/rdkit}
}

@ARTICLE{Matplotlib,
  author={Hunter, John D.},
  journal={Comput. Sci. Eng.},
  title={Matplotlib: A 2D Graphics Environment},
  year={2007},
  volume={9},
  number={3},
  pages={90-95},
  doi={10.1109/MCSE.2007.55}}

@inproceedings{ester1996density,
  title={A density-based algorithm for discovering clusters in large spatial databases with noise.},
  author={Ester, Martin and Kriegel, Hans-Peter and Sander, J{\"o}rg and Xu, Xiaowei and others},
  booktitle={Proceedings of the Second International Conference on Knowledge Discovery and Data Mining},
  volume={96},
  number={34},
  pages={226--231},
  year={1996}
}

@article{schubert2017dbscan,
  title={DBSCAN revisited, revisited: why and how you should (still) use DBSCAN},
  author={Schubert, Erich and Sander, J{\"o}rg and Ester, Martin and Kriegel, Hans Peter and Xu, Xiaowei},
  journal={ACM T. Database Syst.},
  volume={42},
  number={3},
  pages={1--21},
  year={2017},
  publisher={ACM New York, NY, USA}
}

@article{kriegel2011density,
  title={Density-based clustering},
  author={Kriegel, Hans-Peter and Kr{\"o}ger, Peer and Sander, J{\"o}rg and Zimek, Arthur},
  journal={WIREs Data. Min. Knowl.},
  volume={1},
  number={3},
  pages={231--240},
  year={2011},
  publisher={Wiley Online Library}
}

@article{ward1963hierarchical,
  title={Hierarchical grouping to optimize an objective function},
  author={Ward Jr, Joe H},
  journal={J. Am. Stat. Assoc.},
  volume={58},
  number={301},
  pages={236--244},
  year={1963},
  publisher={Taylor \& Francis}
}

@book{kaufman2009finding,
  title={Finding groups in data: an introduction to cluster analysis},
  author={Kaufman, Leonard and Rousseeuw, Peter J},
  year={2009},
  publisher={John Wiley \& Sons}
}

@Article{lloyd1982least,
  author    = {Lloyd, Stuart},
  title     = {Least Squares Quantization in PCM},
  number    = {2},
  pages     = {129--137},
  volume    = {28},
  journal   = {IEEE Trans. Inf. Theory},
  publisher = {IEEE},
  year      = {1982},
}

@inproceedings{macqueen1967some,
  title={Some Methods for Classification and Analysis of Multivariate Observations},
  author={MacQueen, James and others},
  booktitle={Proceedings of the fifth Berkeley symposium on mathematical statistics and probability},
  volume={1},
  number={14},
  pages={281--297},
  year={1967},
  organization={Oakland, CA, USA}
}

@article{ikotunKmeansClusteringAlgorithms2023,
  title = {K-Means Clustering Algorithms: {{A}} Comprehensive Review, Variants Analysis, and Advances in the Era of Big Data},
  author = {Ikotun, Abiodun M. and Ezugwu, Absalom E. and Abualigah, Laith and Abuhaija, Belal and Heming, Jia},
  year = {2023},
  month = apr,
  journal = {Inf. Sci.},
  volume = {622},
  pages = {178--210},
  issn = {00200255},
  doi = {10.1016/j.ins.2022.11.139}
}

@Article{pulay1992geometry,
  author    = {Pulay, Peter and Fogarasi, Geza},
  title     = {Geometry Optimization in Redundant Internal Coordinates},
  number    = {4},
  pages     = {2856--2860},
  volume    = {96},
  journal   = {J. Chem. Phys.},
  publisher = {American Institute of Physics},
  year      = {1992},
}

@Article{peng1996using,
  author    = {Peng, Chunyang and Ayala, Philippe Y and Schlegel, H Bernhard and Frisch, Michael J},
  title     = {Using Redundant Internal Coordinates to Optimize Equilibrium Geometries and Transition States},
  number    = {1},
  pages     = {49--56},
  volume    = {17},
  journal   = {J. Comput. Chem.},
  publisher = {Wiley Online Library},
  year      = {1996},
}

@misc{deepseekai2025deepseekv32pushingfrontieropen,
      title={DeepSeek-V3.2: Pushing the Frontier of Open Large Language Models}, 
      author={DeepSeek-AI and Aixin Liu and Aoxue Mei and Bangcai Lin and Bing Xue and Bingxuan Wang and Bingzheng Xu and Bochao Wu and Bowei Zhang and Chaofan Lin and Chen Dong and Chengda Lu and Chenggang Zhao and Chengqi Deng and Chenhao Xu and Chong Ruan and Damai Dai and Daya Guo and Dejian Yang and Deli Chen and Erhang Li and Fangqi Zhou and Fangyun Lin and Fucong Dai and Guangbo Hao and Guanting Chen and Guowei Li and H. Zhang and Hanwei Xu and Hao Li and Haofen Liang and Haoran Wei and Haowei Zhang and Haowen Luo and Haozhe Ji and Honghui Ding and Hongxuan Tang and Huanqi Cao and Huazuo Gao and Hui Qu and Hui Zeng and Jialiang Huang and Jiashi Li and Jiaxin Xu and Jiewen Hu and Jingchang Chen and Jingting Xiang and Jingyang Yuan and Jingyuan Cheng and Jinhua Zhu and Jun Ran and Junguang Jiang and Junjie Qiu and Junlong Li and Junxiao Song and Kai Dong and Kaige Gao and Kang Guan and Kexin Huang and Kexing Zhou and Kezhao Huang and Kuai Yu and Lean Wang and Lecong Zhang and Lei Wang and Liang Zhao and Liangsheng Yin and Lihua Guo and Lingxiao Luo and Linwang Ma and Litong Wang and Liyue Zhang and M. S. Di and M. Y Xu and Mingchuan Zhang and Minghua Zhang and Minghui Tang and Mingxu Zhou and Panpan Huang and Peixin Cong and Peiyi Wang and Qiancheng Wang and Qihao Zhu and Qingyang Li and Qinyu Chen and Qiushi Du and Ruiling Xu and Ruiqi Ge and Ruisong Zhang and Ruizhe Pan and Runji Wang and Runqiu Yin and Runxin Xu and Ruomeng Shen and Ruoyu Zhang and S. H. Liu and Shanghao Lu and Shangyan Zhou and Shanhuang Chen and Shaofei Cai and Shaoyuan Chen and Shengding Hu and Shengyu Liu and Shiqiang Hu and Shirong Ma and Shiyu Wang and Shuiping Yu and Shunfeng Zhou and Shuting Pan and Songyang Zhou and Tao Ni and Tao Yun and Tian Pei and Tian Ye and Tianyuan Yue and Wangding Zeng and Wen Liu and Wenfeng Liang and Wenjie Pang and Wenjing Luo and Wenjun Gao and Wentao Zhang and Xi Gao and Xiangwen Wang and Xiao Bi and Xiaodong Liu and Xiaohan Wang and Xiaokang Chen and Xiaokang Zhang and Xiaotao Nie and Xin Cheng and Xin Liu and Xin Xie and Xingchao Liu and Xingkai Yu and Xingyou Li and Xinyu Yang and Xinyuan Li and Xu Chen and Xuecheng Su and Xuehai Pan and Xuheng Lin and Xuwei Fu and Y. Q. Wang and Yang Zhang and Yanhong Xu and Yanru Ma and Yao Li and Yao Li and Yao Zhao and Yaofeng Sun and Yaohui Wang and Yi Qian and Yi Yu and Yichao Zhang and Yifan Ding and Yifan Shi and Yiliang Xiong and Ying He and Ying Zhou and Yinmin Zhong and Yishi Piao and Yisong Wang and Yixiao Chen and Yixuan Tan and Yixuan Wei and Yiyang Ma and Yiyuan Liu and Yonglun Yang and Yongqiang Guo and Yongtong Wu and Yu Wu and Yuan Cheng and Yuan Ou and Yuanfan Xu and Yuduan Wang and Yue Gong and Yuhan Wu and Yuheng Zou and Yukun Li and Yunfan Xiong and Yuxiang Luo and Yuxiang You and Yuxuan Liu and Yuyang Zhou and Z. F. Wu and Z. Z. Ren and Zehua Zhao and Zehui Ren and Zhangli Sha and Zhe Fu and Zhean Xu and Zhenda Xie and Zhengyan Zhang and Zhewen Hao and Zhibin Gou and Zhicheng Ma and Zhigang Yan and Zhihong Shao and Zhixian Huang and Zhiyu Wu and Zhuoshu Li and Zhuping Zhang and Zian Xu and Zihao Wang and Zihui Gu and Zijia Zhu and Zilin Li and Zipeng Zhang and Ziwei Xie and Ziyi Gao and Zizheng Pan and Zongqing Yao and Bei Feng and Hui Li and J. L. Cai and Jiaqi Ni and Lei Xu and Meng Li and Ning Tian and R. J. Chen and R. L. Jin and S. S. Li and Shuang Zhou and Tianyu Sun and X. Q. Li and Xiangyue Jin and Xiaojin Shen and Xiaosha Chen and Xinnan Song and Xinyi Zhou and Y. X. Zhu and Yanping Huang and Yaohui Li and Yi Zheng and Yuchen Zhu and Yunxian Ma and Zhen Huang and Zhipeng Xu and Zhongyu Zhang and Dongjie Ji and Jian Liang and Jianzhong Guo and Jin Chen and Leyi Xia and Miaojun Wang and Mingming Li and Peng Zhang and Ruyi Chen and Shangmian Sun and Shaoqing Wu and Shengfeng Ye and T. Wang and W. L. Xiao and Wei An and Xianzu Wang and Xiaowen Sun and Xiaoxiang Wang and Ying Tang and Yukun Zha and Zekai Zhang and Zhe Ju and Zhen Zhang and Zihua Qu},
      year={2025},
      eprint={2512.02556},
      archivePrefix={arXiv},
      primaryClass={cs.CL},
      url={https://arxiv.org/abs/2512.02556}, 
}

@misc{bytedance2025doubaoseed,
  author       = {{ByteDance}},
  title        = {{Doubao-Seed-1.6-Vision}: A Multimodal Large Language Model},
  year         = {2025},
  note         = {Accessed: September 30, 2025},
  howpublished = {VolcEngine / ByteDance AI},
  url          = {https://www.volcengine.com/product/doubao}
}

@article{tomar2006converting,
  title={Converting video formats with FFmpeg},
  author={Tomar, Suramya},
  journal={Linux J.},
  volume={2006},
  number={146},
  pages={10},
  year={2006},
  publisher={Belltown Media}
}

@misc{liu2025automatedframeworkanalyzingstructural,
      title={An Automated Framework for Analyzing Structural Evolution in On-the-fly Non-adiabatic Molecular Dynamics Using Autoencoder and Multiple Molecular Descriptors}, 
      author={Hangxu Liu and Yifei Zhu and Zhenggang Lan},
      year={2025},
      eprint={2511.13364},
      archivePrefix={arXiv},
      primaryClass={physics.chem-ph},
      url={https://arxiv.org/abs/2511.13364}, 
}

@book{domcke2004conical,
  title={Conical Intersections: Electronic Structure, Dynamics \& Spectroscopy},
  author={Domcke, Wolfgang and Yarkony, David and K{\"o}ppel, Horst},
  volume={15},
  year={2004},
  publisher={World Scientific}
}

@Article{domcke2012role,
  author    = {Domcke, Wolfgang and Yarkony, David R},
  title     = {Role of Conical Intersections in Molecular Spectroscopy and Photoinduced Chemical Dynamics},
  pages     = {325--352},
  volume    = {63},
  journal   = {Annu. Rev. Phys. Chem},
  publisher = {Annual Reviews},
  year      = {2012},
}

@Book{domcke2011conical,
  author    = {Domcke, Wolfgang and Yarkony, David R and K{\"o}ppel, Horst},
  title     = {Conical Intersections: Theory, Computation, and Experiment},
  publisher = {World Scientific},
  volume    = {17},
  year      = {2011},
}

@Article{matsika2011nonadiabatic,
  author    = {Matsika, Spiridoula and Krause, Pascal},
  title     = {Nonadiabatic Events and Conical Intersections},
  pages     = {621--643},
  volume    = {62},
  journal   = {Annu. Rev. Phys. Chem},
  publisher = {Annual Reviews},
  year      = {2011},
}

@article{matsika2021electronic,
  title={Electronic structure methods for the description of nonadiabatic effects and conical intersections},
  author={Matsika, Spiridoula},
  journal={Chem. Rev.},
  volume={121},
  number={15},
  pages={9407--9449},
  year={2021},
  publisher={ACS Publications}
}

@Article{crespo2018recent,
  author    = {{Crespo-Otero}, Rachel and Barbatti, Mario},
  title     = {Recent Advances and Perspectives on Nonadiabatic Mixed Quantum--Classical Dynamics},
  doi       = {10.1021/acs.chemrev.7b00577},
  number    = {15},
  pages     = {7026--7068},
  volume    = {118},
  journal   = {Chem. Rev.},
  publisher = {ACS Publications},
  year      = {2018},
}

@inbook{Gonzalez2020book,
author = {G{\'o}mez, Sandra and Galván, Ignacio Fdez. and Lindh, Roland and Gonz{\'a}lez, Leticia},
publisher = {John Wiley \& Sons, Ltd},
address = {USA},
isbn = {9781119417774},
title = {Motivation and Basic Concepts},
booktitle = {Quantum Chemistry and Dynamics of Excited States},
chapter = {1},
pages = {1-12},
editor={L. Gonz{\'a}lez and R. Lindh},
year = {2020},
}

@Article{curchod2018ab,
  author    = {Curchod, Basile FE and Mart{\'{\i}}nez, Todd J},
  title     = {Ab Initio Nonadiabatic Quantum Molecular Dynamics},
  doi       = {10.1021/acs.chemrev.7b00423},
  number    = {7},
  pages     = {3305--3336},
  volume    = {118},
  journal   = {Chem. Rev.},
  publisher = {ACS Publications},
  year      = {2018},
}

@article{mai2018nonadiabatic,
  title={Nonadiabatic dynamics: The SHARC approach},
  author={Mai, Sebastian and Marquetand, Philipp and Gonz{\'a}lez, Leticia},
  journal={WIREs Comput. Mol. Sci.},
  volume={8},
  number={6},
  pages={e1370},
  year={2018},
  publisher={Wiley Online Library}
}

@article{akimov2013theoretical,
  title={Theoretical insights into photoinduced charge transfer and catalysis at oxide interfaces},
  author={Akimov, Alexey V and Neukirch, Amanda J and Prezhdo, Oleg V},
  journal={Chem. Rev.},
  volume={113},
  number={6},
  pages={4496--4565},
  year={2013},
  publisher={ACS Publications}
}

@article{du2015fly,
  title={An on-the-fly surface-hopping program jade for nonadiabatic molecular dynamics of polyatomic systems: implementation and applications},
  author={Du, Likai and Lan, Zhenggang},
  journal={J. Chem. Theory Comput.},
  volume={11},
  number={4},
  pages={1360--1374},
  year={2015},
  publisher={ACS Publications}
}

@article{tapavicza2007trajectory,
  title={Trajectory surface hopping within linear response time-dependent density-functional theory},
  author={Tapavicza, Enrico and Tavernelli, Ivano and Rothlisberger, Ursula},
  journal={Phys. Rev. Lett.},
  volume={98},
  number={2},
  pages={023001},
  year={2007},
  publisher={APS}
}

@article{nelson2014nonadiabatic,
  title={Nonadiabatic excited-state molecular dynamics: Modeling photophysics in organic conjugated materials},
  author={Nelson, Tammie and Fernandez-Alberti, Sebastian and Roitberg, Adrian E and Tretiak, Sergei},
  journal={Acc. Chem. Res.},
  volume={47},
  number={4},
  pages={1155--1164},
  year={2014},
  publisher={ACS Publications}
}

@article{tully2012perspective,
  title={Perspective: Nonadiabatic dynamics theory},
  author={Tully, John C},
  journal={J. Chem. Phys.},
  volume={137},
  number={22},
  pages={22A301},
  year={2012},
  publisher={American Institute of Physics}
}

@article{virshup2012nonlinear,
  title = {Nonlinear Dimensionality Reduction for Nonadiabatic Dynamics: {{The}} Influence of Conical Intersection Topography on Population Transfer Rates},
  author = {Virshup, Aaron M and Chen, Jiahao and Mart{\'i}nez, Todd J},
  year = 2012,
  journal = {J. Chem. Phys.},
  volume = {137},
  number = {22},
  pages = {22A519},
  publisher = {American Institute of Physics},
  doi = {10.1063/1.4742066},
  copyright = {3.6083},
  lccn = {4.3037}
}

@article{ceriottiUnsupervisedMachineLearning2019,
  title = {Unsupervised Machine Learning in Atomistic Simulations, between Predictions and Understanding},
  author = {Ceriotti, Michele},
  year = 2019,
  month = apr,
  journal = {J. Chem. Phys.},
  volume = {150},
  number = {15},
  pages = {150901},
  publisher = {American Institute of Physics},
  issn = {0021-9606},
  doi = {10.1063/1.5091842}
}

@article{glielmo2021unsupervised,
  title = {Unsupervised Learning Methods for Molecular Simulation Data},
  author = {Glielmo, Aldo and Husic, Brooke E and Rodriguez, Alex and Clementi, Cecilia and No{\'e}, Frank and Laio, Alessandro},
  year = 2021,
  journal = {Chem. Rev.},
  volume = {121},
  number = {16},
  pages = {9722--9758},
  publisher = {ACS Publications},
  doi = {10.1021/acs.chemrev.0c01195}
}

@misc{li2025chemvlmexploringpowermultimodal,
      title={ChemVLM: Exploring the Power of Multimodal Large Language Models in Chemistry Area}, 
      author={Junxian Li and Di Zhang and Xunzhi Wang and Zeying Hao and Jingdi Lei and Qian Tan and Cai Zhou and Wei Liu and Yaotian Yang and Xinrui Xiong and Weiyun Wang and Zhe Chen and Wenhai Wang and Wei Li and Shufei Zhang and Mao Su and Wanli Ouyang and Yuqiang Li and Dongzhan Zhou},
      year={2025},
      eprint={2408.07246},
      archivePrefix={arXiv},
      primaryClass={cs.LG},
      url={https://arxiv.org/abs/2408.07246}, 
}

@article{chenLargescaleChemicalReaction2025,
  title = {Towards Large-Scale Chemical Reaction Image Parsing {\emph{via}} a Multimodal Large Language Model},
  author = {Chen, Yufan and Leung, Ching Ting and Sun, Jianwei and Huang, Yong and Li, Linyan and Chen, Hao and Gao, Hanyu},
  year = 2025,
  journal = {Chem. Sci.},
  volume = {16},
  number = {45},
  pages = {21464--21474},
  issn = {2041-6520, 2041-6539},
  doi = {10.1039/D5SC04173B}
}

@misc{ai4science2023impactlargelanguagemodels,
      title={The Impact of Large Language Models on Scientific Discovery: a Preliminary Study using GPT-4}, 
      author={Microsoft Research AI4Science and Microsoft Azure Quantum},
      year={2023},
      eprint={2311.07361},
      archivePrefix={arXiv},
      primaryClass={cs.CL},
      url={https://arxiv.org/abs/2311.07361}, 
}

@inproceedings{
jimenez2024swebench,
title={{SWE}-bench: Can Language Models Resolve Real-world Github Issues?},
author={Carlos E Jimenez and John Yang and Alexander Wettig and Shunyu Yao and Kexin Pei and Ofir Press and Karthik R Narasimhan},
booktitle={The Twelfth International Conference on Learning Representations},
year={2024},
url={https://openreview.net/forum?id=VTF8yNQM66}
}

@misc{laurent2024labbenchmeasuringcapabilitieslanguage,
      title={LAB-Bench: Measuring Capabilities of Language Models for Biology Research}, 
      author={Jon M. Laurent and Joseph D. Janizek and Michael Ruzo and Michaela M. Hinks and Michael J. Hammerling and Siddharth Narayanan and Manvitha Ponnapati and Andrew D. White and Samuel G. Rodriques},
      year={2024},
      eprint={2407.10362},
      archivePrefix={arXiv},
      primaryClass={cs.AI},
      url={https://arxiv.org/abs/2407.10362}, 
}

@article{miret2025enabling,
  title={Enabling large language models for real-world materials discovery},
  author={Miret, Santiago and Krishnan, NM Anoop},
  journal={Nat. Mach. Intell.},
  volume={7},
  number={7},
  pages={991--998},
  year={2025},
  publisher={Nature Publishing Group UK London}
}

@article{white2023future,
  title={The future of chemistry is language},
  author={White, Andrew D},
  journal={Nat. Rev. Chem.},
  volume={7},
  number={7},
  pages={457--458},
  year={2023},
  publisher={Nature Publishing Group UK London}
}

@article{jablonka202314,
  title={14 examples of how LLMs can transform materials science and chemistry: a reflection on a large language model hackathon},
  author={Jablonka, Kevin Maik and Ai, Qianxiang and Al-Feghali, Alexander and Badhwar, Shruti and Bocarsly, Joshua D and Bran, Andres M and Bringuier, Stefan and Brinson, L Catherine and Choudhary, Kamal and Circi, Defne and others},
  journal={Digit. Discov.},
  volume={2},
  number={5},
  pages={1233--1250},
  year={2023},
  publisher={Royal Society of Chemistry}
}

@article{ramos2025review,
  title={A review of large language models and autonomous agents in chemistry},
  author={Ramos, Mayk Caldas and Collison, Christopher J and White, Andrew D},
  journal={Chem. Sci.},
  year={2025},
  publisher={Royal Society of Chemistry}
}

@article{chenSciToolAgentKnowledgeGraphDriven2025,
  title = {{{SciToolAgent}}: {{A Knowledge Graph-Driven Scientific Agent}} for {{Multi-Tool Integration}}},
  shorttitle = {{{SciToolAgent}}},
  author = {Chen, Huajun and Ding, Keyan and Yu, Jing and Huang, Junjie and Yang, Yuchen and Zhang, Qiang},
  year = 2025,
  month = jan,
  journal = {Nat. Comput. Sci.},
  doi = {10.21203/rs.3.rs-5610718/v1},
  copyright = {https://creativecommons.org/licenses/by/4.0/}
}

@Article{abdi2010principal,
  author    = {Abdi, Herv{\'e} and Williams, Lynne J},
  title     = {Principal Component Analysis},
  doi       = {10.1002/wics.101},
  number    = {4},
  pages     = {433--459},
  volume    = {2},
  journal   = {WIREs Comput. Stat.},
  publisher = {{Wiley Online Library}},
  year      = {2010},
}

@article{wold1987principal,
author = {Wold, Svante and Esbensen, Kim and Geladi, Paul},
year = {1987},
month = {08},
pages = {37-52},
title = {Principal Component Analysis},
volume = {2},
journal = {Chemometr. Intell. Lab.},
doi = {10.1016/0169-7439(87)80084-9}
}

@article{achesonAutomaticClusteringExcitedState2023,
  title = {Automatic {{Clustering}} of {{Excited-State Trajectories}}: {{Application}} to {{Photoexcited Dynamics}}},
  shorttitle = {Automatic {{Clustering}} of {{Excited-State Trajectories}}},
  author = {Acheson, Kyle and Kirrander, Adam},
  year = 2023,
  month = sep,
  journal = {J. Chem. Theory Comput.},
  volume = {19},
  number = {18},
  pages = {6126--6138},
  issn = {1549-9618, 1549-9626},
  doi = {10.1021/acs.jctc.3c00776},
  copyright = {https://creativecommons.org/licenses/by/4.0/}
}

@article{belyaev2015nonadiabatic,
  title = {Nonadiabatic Nuclear Dynamics of the Ammonia Cation Studied by Surface Hopping Classical Trajectory Calculations},
  author = {Belyaev, Andrey K and Domcke, Wolfgang and Lasser, Caroline and Trigila, Giulio},
  year = 2015,
  journal = {J. Chem. Phys.},
  volume = {142},
  number = {10},
  pages = {104307},
  publisher = {AIP Publishing LLC},
  copyright = {3.6083},
  lccn = {4.3037}
}

@article{capano2017photophysics,
  title = {Photophysics of a Copper Phenanthroline Eluciyeard by Trajectory and Wavepacket-Based Quantum Dynamics: A Synergetic Approach},
  author = {Capano, G and Penfold, {\relax TJ} and Chergui, M and Tavernelli, I},
  year = 2017,
  journal = {Phys. Chem. Chem. Phys.},
  volume = {19},
  number = {30},
  pages = {19590--19600},
  publisher = {Royal Society of Chemistry},
  doi = {10.1039/C7CP00436B},
  copyright = {3.8610},
  lccn = {3.9451}
}

@article{delmasAutomatedSelectionNuclear2025,
  title = {Automated {{Selection}} of {{Nuclear Coordinates}} for {{Reduced Dimensionality Nonadiabatic Dynamics}}},
  author = {Delmas, Vincent and Nardi, Alessandro Nicola and Merritt, Isabella C. D. and Fert{\'e}, Anthony and Fdez. Galv{\'a}n, Ignacio and Vacher, Morgane},
  year = 2025,
  month = jul,
  journal = {J. Chem. Theory Comput.},
  volume = {21},
  number = {13},
  pages = {6611--6621},
  issn = {1549-9618, 1549-9626},
  doi = {10.1021/acs.jctc.5c00110},
  copyright = {https://doi.org/10.15223/policy-029}
}

@article{how2021significance,
  title = {Significance of the Chemical Environment of an Element in Nonadiabatic Molecular Dynamics: {{Feature}} Selection and Dimensionality Reduction with Machine Learning},
  author = {How, Wei Bin and Wang, Bipeng and Chu, Weibin and Tkatchenko, Alexandre and Prezhdo, Oleg V},
  year = 2021,
  journal = {J. Phys. Chem. Lett.},
  volume = {12},
  number = {50},
  pages = {12026--12032},
  publisher = {ACS Publications},
  doi = {10.1021/acs.jpclett.1c03469},
  copyright = {7.301},
  lccn = {6.888}
}

@article{how2022dimensionality,
  title = {Dimensionality Reduction in Machine Learning for Nonadiabatic Molecular Dynamics: {{Effectiveness}} of Elemental Sublattices in Lead Halide Perovskites},
  author = {How, Wei Bin and Wang, Bipeng and Chu, Weibin and Kovalenko, Sergiy M and Tkatchenko, Alexandre and Prezhdo, Oleg V},
  year = 2022,
  journal = {J. Chem. Phys.},
  volume = {156},
  number = {5},
  pages = {054110},
  publisher = {AIP Publishing LLC},
  doi = {10.1063/5.0078473}
}

@article{karaDONKEYFlexibleAccurate2025,
  title = {{{DONKEY}}: {{A Flexible}} and {{Accurate Algorithm}} for {{Clustering}}},
  shorttitle = {{{DONKEY}}},
  author = {K{\'a}ra, Jakub and Acheson, Kyle and Kirrander, Adam},
  year = 2025,
  month = may,
  journal = {J. Chem. Theory Comput.},
  pages = {acs.jctc.4c01750},
  issn = {1549-9618, 1549-9626},
  doi = {10.1021/acs.jctc.4c01750},
  copyright = {https://creativecommons.org/licenses/by/4.0/}
}

@article{kochmanNonadiabaticMolecularDynamics2024,
  title = {Nonadiabatic Molecular Dynamics Simulations Shed Light on the Timescale of Furylfulgide Photocyclisation},
  author = {Kochman, Micha{\l} Andrzej},
  year = 2024,
  journal = {New J. Chem.},
  volume = {48},
  number = {32},
  pages = {14327--14335},
  issn = {1144-0546, 1369-9261},
  doi = {10.1039/D3NJ04752K}
}

@article{liAnalysisGeometricalEvolution2017,
  title = {Analysis of the {{Geometrical Evolution}} in {{On-the-Fly Surface-Hopping Nonadiabatic Dynamics}} with {{Machine Learning Dimensionality Reduction Approaches}}: {{Classical Multidimensional Scaling}} and {{Isometric Feature Mapping}}},
  shorttitle = {Analysis of the {{Geometrical Evolution}} in {{On-the-Fly Surface-Hopping Nonadiabatic Dynamics}} with {{Machine Learning Dimensionality Reduction Approaches}}},
  author = {Li, Xusong and Xie, Yu and Hu, Deping and Lan, Zhenggang},
  year = 2017,
  month = oct,
  journal = {J. Chem. Theory Comput.},
  volume = {13},
  number = {10},
  pages = {4611--4623},
  issn = {1549-9618, 1549-9626},
  doi = {10.1021/acs.jctc.7b00394},
  copyright = {6.440},
  lccn = {6.578}
}

@article{mangan2021dependence,
  title = {Dependence between Structural and Electronic Properties of {{CsPbI3}}: {{Unsupervised}} Machine Learning of Nonadiabatic Molecular Dynamics},
  author = {Mangan, Spencer M and Zhou, Guoqing and Chu, Weibin and Prezhdo, Oleg V},
  year = 2021,
  journal = {J. Phys. Chem. Lett.},
  volume = {12},
  number = {35},
  pages = {8672--8678},
  publisher = {ACS Publications},
  doi = {10.1021/acs.jpclett.1c02361},
  copyright = {7.301},
  lccn = {6.888}
}

@article{pengAnalysisBathMotion2021a,
  title = {Analysis of Bath Motion in {{MM-SQC}} Dynamics via Dimensionality Reduction Approach: {{Principal}} Component Analysis},
  shorttitle = {Analysis of Bath Motion in {{MM-SQC}} Dynamics via Dimensionality Reduction Approach},
  author = {Peng, Jiawei and Xie, Yu and Hu, Deping and Lan, Zhenggang},
  year = 2021,
  month = mar,
  journal = {J. Chem. Phys.},
  volume = {154},
  number = {9},
  pages = {094122},
  issn = {0021-9606, 1089-7690},
  doi = {10.1063/5.0039743}
}

@article{richings2021analyzing,
  title = {Analyzing Grid-Based Direct Quantum Molecular Dynamics Using Non-Linear Dimensionality Reduction},
  author = {Richings, Gareth W and Habershon, Scott},
  year = 2021,
  journal = {Molecules},
  volume = {26},
  number = {24},
  pages = {7418},
  publisher = {MDPI},
  doi = {10.3390/molecules26247418}
}

@article{tavadze2018machine,
  title = {A Machine-Driven Hunt for Global Reaction Coordinates of Azobenzene Photoisomerization},
  author = {Tavadze, Pedram and Avenda{\~n}o Franco, Guillermo and Ren, Pengju and Wen, Xiaodong and Li, Yongwang and Lewis, James P},
  year = 2018,
  journal = {J. Am. Chem. Soc.},
  volume = {140},
  number = {1},
  pages = {285--290},
  publisher = {ACS Publications},
  doi = {10.1021/jacs.7b10030},
  copyright = {16.289},
  lccn = {16.383}
}

@article{zhou2020structural,
  title = {Structural Deformation Controls Charge Losses in {{MAPbI3}}: {{Unsupervised}} Machine Learning of Nonadiabatic Molecular Dynamics},
  author = {Zhou, Guoqing and Chu, Weibin and Prezhdo, Oleg V},
  year = 2020,
  journal = {ACS Energy Lett.},
  volume = {5},
  number = {6},
  pages = {1930--1938},
  publisher = {ACS Publications},
  doi = {10.1021/acsenergylett.0c00899},
  copyright = {24.052},
  lccn = {23.991}
}

@article{tsutsumiReactionSpaceProjector2022,
  title = {Reaction {{Space Projector}} ({{ReSPer}}) for {{Visualizing Dynamic Reaction Routes Based}} on {{Reduced-Dimension Space}}},
  author = {Tsutsumi, Takuro and Ono, Yuriko and Taketsugu, Tetsuya},
  year = 2022,
  month = may,
  journal = {Top. Curr. Chem.},
  volume = {380},
  number = {3},
  pages = {19},
  issn = {2365-0869, 2364-8961},
  doi = {10.1007/s41061-022-00377-7}
}

@article{tsutsumiVisualizationIntrinsicReaction2018,
  title = {Visualization of the {{Intrinsic Reaction Coordinate}} and {{Global Reaction Route Map}} by {{Classical Multidimensional Scaling}}},
  author = {Tsutsumi, Takuro and Ono, Yuriko and Arai, Zin and Taketsugu, Tetsuya},
  year = 2018,
  month = aug,
  journal = {J. Chem. Theory Comput.},
  volume = {14},
  number = {8},
  pages = {4263--4270},
  issn = {1549-9618, 1549-9626},
  doi = {10.1021/acs.jctc.8b00176}
}
\end{document}